\documentclass[10pt]{article}
\usepackage[utf8]{inputenc}

\usepackage{setspace}

\usepackage{tikz}
\definecolor{colorpos}{HTML}{006400}
\definecolor{colorneg}{HTML}{FF0000}

\usepackage[unicode,linktoc=all,hidelinks]{hyperref}
\usepackage[a4paper, margin=2cm]{geometry}
\usepackage{mathtools}

\usepackage[T1]{fontenc}
\usepackage{calligra,amsmath,amssymb}

\newcommand{\w}{\text{\calligra w}\,}
\newcommand{\bb}{\text{\calligra b}\,}
\newcommand{\s}{\text{\calligra s}\,}

\usepackage{amsmath}
\usepackage{amssymb}
\usepackage{amsthm}
\usepackage{bm}
\usepackage{bbm}

\newcommand{\indep}{\perp \!\!\! \perp}

\DeclareSymbolFont{cmletters}{OT1}{cmr}{m}{n}
\DeclareMathSymbol{\Ups}{\mathalpha}{cmletters}{"7}

\usepackage{float} 
\usepackage{adjustbox}
\usepackage{threeparttable}
\usepackage{booktabs}

\usepackage{algorithm}
\usepackage{algorithmic}

\usepackage[english]{babel}
\usepackage{natbib}

\usepackage{graphicx}

\usepackage[title]{appendix}

\newtheorem{definition}{Definition}
\newtheorem{assumption}{Assumption}
\newtheorem{theorem}{Theorem}
\newtheorem{corollary}{Corollary}

\usepackage{algorithm}
\usepackage{algorithmic}

\title{A Spatial Autoregressive Graphical Model}
\author{Sjoerd Hermes$^{1,2}$, Joost van Heerwaarden$^{1,2}$ and Pariya Behrouzi$^1$}
\date{%
    $^1$ Mathematical and Statistical Methods, Wageningen University\\%
    $^2$ Plant Production Systems, Wageningen University\\
}

\begin{document}

\maketitle
\begin{abstract}
\noindent Within the statistical literature, a significant gap exists in methods capable of modeling asymmetric multivariate spatial effects that elucidate the relationships underlying complex spatial phenomena. For such a phenomenon, observations at any location are expected to arise from a combination of within- and between-location effects, where the latter exhibit asymmetry. This asymmetry is represented by heterogeneous spatial effects between locations pertaining to different categories, that is, a feature inherent to each location in the data, such that based on the feature label, asymmetric spatial relations are postulated between neighbouring locations with different labels. Our novel approach synergises the principles of multivariate spatial autoregressive models and the Gaussian graphical model. This synergy enables us to effectively address the gap by accommodating asymmetric spatial relations, overcoming the usual constraints in spatial analyses. Using a Bayesian-estimation framework, the model performance is assessed in a simulation study. We apply the model on intercropping data, where spatial effects between different crops are unlikely to be symmetric, in order to illustrate the usage of the proposed methodology. An R package containing the proposed methodology can be found on \url{https://CRAN.R-project.org/package=SAGM}.
\end{abstract}

\noindent%

\section{Introduction}
Gaussian graphical models are statistical learning techniques used to make inference on conditional dependence relationships within a set of variables arising from a multivariate normal distribution (Lauritzen, \citeyear{lauritzen1996graphical}). Methodological developments have expanded the use of the Gaussian graphical model beyond non-temporal to temporal data, through, for example, the use of time-varying or autoregressive graph structures (see for relevant research during the last 5 years Barigozzi \& Brownlees, \citeyear{barigozzi2019nets}; Paci \& Consonni, \citeyear{paci2020structural}; Yang \& Peng, \citeyear{yang2020estimating}; Dallakyan et al., \citeyear{dallakyan2022time}). These developments allow for improved analyses of complex, multivariate temporal datasets. However, barely any attention has been given to the development of graphical models pertaining to the higher dimensional cousin of time: space. Where dynamic models deal with temporal variation, assuming data close in time (Yang \& Peng, \citeyear{yang2020estimating}), there are clear parallels with the spatial analogue, for which data close together in space can be expected to be similar due to potential presence of location-specific latent factors affecting the distributions of the data that pertain to the location. 

One example of spatial dependence between different variables can be found in intercropping trials. Intercropping is the cultivation of  multiple crop species in a single field. Certain traits of the crops, such as yield, are not only dependent on within-crop processes, but also depend on the effects of traits expressed by different crops growing on neighbouring plots (subsections of fields that contain multiple plants of the same crop). These effects can be either  positive (complementarity and facilitation) or negative (competition) (Bourke et al., \citeyear{bourke2021breeding}) with respect to biomass production and yield. Knowledge of such effects is invaluable in selecting intercropping combinations with the goal of yield-maximisation (Bourke et al., \citeyear{bourke2021breeding}). Contemporary research on intercropping systems has shown that these spatial effects exhibit asymmetry (Huang et al., \citeyear{huang2017plant}; Gao et al., \citeyear{gao2021common}), that is, the effect of variables pertaining to crop $c_1$ on variables pertaining to crop $c_2$ is not the same as the effect of those same variables pertaining to crop $c_2$ on those of crop $c_1$.
\\
\\
Despite a growing interest in the design of productive intercropping systems (Federer \citeyear{federer2012statistical}), there has been little methodological development around the identification of the kind of multi-trait and multi-species interactions that would determine which crops should ideally be combined (Brooker et al., \citeyear{brooker2015improving}). Given that intercropping data typically consist of observations on locations in a finite domain, the set of spatial autoregressive models (Ord, \citeyear{ord1975estimation}) makes for a logical starting point in this methodological pursuit. An application of this model on yield data in a monocropping system can be found in Long (\citeyear{long1998spatial}). Whilst being suitable for data arising from monocropping systems, this approach is unsuitable for modern intercropping type data that consist of multiple traits measured across multiple crops, as the spatial autoregressive model is univariate in its response and has no way of isolating asymmetric spatial effects or within-plot effects. Another example of an existing approach, by Dobra (\citeyear{dobra2011bayesian}; \citeyear{dobra2016graphical}), explicitly accounts for spatial autocorrelation by developing Bayesian models that construct two graphs: a neighbourhood graph where the vertices indicate different regions and the presence of an edge indicates whether the regions share a border (are neighbours), and a conditional dependence graph that shows which variables are independent given all other variables. However, this approach has no way of inferring the spatial effects of one variable of crop $c_1$ on another of crop $c_2$, which is of interest to researchers wanting to evaluate, for example, the impact of applying fertiliser on one plot on the growth of plants in surrounding plots. Whilst other spatial (autoregressive) models exist (Ord, \citeyear{ord1975estimation}; Yang \& Lee, \citeyear{yang2017identification}), none can model heterogeneous multivariate spatial effects across multiple crops, where the heterogeneity arises from differing spatial effects per crop, whilst simultaneously capturing the complex dependence relations that occur within plots. 
\\
\\
The need for such a method extends beyond intercropping applications. Any phenomenon in which asymmetric spatial effects between multiple categories can be postulated provides a potential application. The categories are features inherent to each location in the data, such that based on the label of that feature, asymmetric spatial relations can occur between neighbouring locations with different feature labels. One potential area of application is the field of epidemiology, where researchers might want to evaluate the existence and extend of asymmetric effects in rural-urban disease transmissions and driving variables (Ferrari et al., \citeyear{ferrari2010rural}). Alternatively, for a more timely example, whether there is asymmetry among the inter-country transmission rates of COVID-19 between neighbouring countries, where the countries can be categorised in strict anti-COVID policy --- assuming that the movement of people between countries remains possible --- versus lax anti-COVID policy (Keita, \citeyear{keita2020air}). A last example is within economics or sociology, where observed values on variables such as crime, unemployment rate, income per capita, population size and congestion within a city are not only a result of within-city processes, but are also affected by what happens in neighbouring cities (Goetzke, \citeyear{goetzke2008network}). The asymmetry arises, for example, due to different spatial relations between cities in the USA with a democratic versus a republication major, which in turn might imply different policies being put into practice that affect the aforementioned variables. 
\\
\\
In line with Dahlhaus and Eichler (\citeyear{dahlhaus2003causality}), who proposed a time series (vector autoregressive) chain graph, showcasing contemporaneous conditional dependencies and dynamic effects, we propose a spatial autoregressive graphical model that fills the methodological gap of methods than can capture asymmetric between-location spatial effects together with within-location effects. Our proposed spatial autoregressive graphical model builds on the recently proposed multivariate spatial autoregressive model (MSAR) by Yang and Lee (\citeyear{yang2017identification}), who extended the univariate spatial autoregressive model (SAR) to the multivariate setting. The asymmetry of spatial effects between different categories, i.e.\ crops in intercropping, is captured through a straightforward manipulation of the spatial weight matrices of the MSAR model, whilst the within-location dependencies share some of the properties of the Gaussian graphical model; namely that the within-location effects are derived from an underlying network consisting of conditional dependencies, which, in turn, offer a parsimonious representation of the complex within-location dependencies as well as being easily interpretable for researchers. The proposed method has attractive asymmetric between-location independence relationships, that are rarely present in spatial autoregressive models. Using this approach, we can identify positive and negative interaction effects between crops that optimise collective performance, thereby selecting combinations of genotypes or crops that are promising in intercropping situations. 

We propose a Bayesian estimation framework and introduce an R package such that researchers can analyse their own spatial data. The R package can be found at \url{https://CRAN.R-project.org/package=SAGM}.
\\
\\
This article proposes a new statistical methodology: the spatial autoregressive graphical model. The methodological novelty arises from the method's capacity to learn multivariate asymmetric between-location effects, combined with the capacity of illustrating complex within-location effects through a conditional independence structure, whereby the between- and within-location effects are illustrated by means of a graph, thereby facilitating interpretability. Section \ref{Methodology}, introduces and formalises the methodological framework. Bayesian inference is discussed in Section \ref{Bayesian inference}. Section \ref{Simulation study} presents an elaborate simulation study, where the performance of the newly proposed method is evaluated on simulated spatial data. An application of the new method on real intercropping data, illustrating the usage of the proposed methodology, is given in Section \ref{Real world data example}. Finally, the conclusion and discussion can be found in Section \ref{Conclusion}. 

\section{Methodology} \label{Methodology}
\subsection{Spatial autoregressive models}
The proposed methodology requires some background knowledge about the MSAR model. Assume data arising from a stochastic process $\{X(i): i = 1,\ldots,n\}$, with $\bigcup_{i = 1}^n i = \mathcal{D}$ and $i \cap i' = \emptyset$ for all $i \neq i'$. Ergo: the data are lattice data, where the lattice $\mathcal{D}$ is a non-random and finite domain. The MSAR (Yang \& Lee, \citeyear{yang2017identification}) is given by the following equation
\begin{equation} \label{eq:msar}
    \bm{X} = \bm{WX}\bm{\Psi} + \bm{E},
\end{equation}
with $\bm{X} \in \mathbb{R}^{n \times p}, \bm{W} \in [0,1]^{n \times n}$, $\bm{\Psi} \in \mathbb{R}^{p \times p}$ and $\bm{E} \in \mathbb{R}^{n \times p}$ and where the row vectors of $\bm{E} = (\bm{E}_{1},\ldots,\bm{E}_{n})^T, \bm{E}_i = (\varepsilon_{i1},\ldots,\varepsilon_{ip}), 1 \leq i \leq n$, are assumed to be i.i.d.\ normally distributed with mean vector $\bm{0}$ and positive definite covariance matrix $ \bm{\Sigma}_E$. $\bm{W}$ is a known spatial weight matrix and spatial effect matrix $\bm{\Psi}$ indicates how variables in one location affects the value of variables in neighbouring locations and vice versa. For the MSAR, the spatial effects captured by $\bm{\Psi}$ remain the same across all locations in $\mathcal{D}$.

To understand the concept of spatial weight matrices, an introduction to the adjacency matrix $\bm{A}$ is necessary, as $\bm{A}$ determines the structure of the spatial weight matrix $\bm{W}$. Element $a_{ij} = a_{ji} = 1,  1 \leq i \neq j \leq n$, if location $i$ is neighboured by location $j$ and 0 otherwise. Moreover, $a_{ii} = 0$ for all $i$. $\bm{W}$ is a row-normalised version of $\bm{A}$, i.e.\ $w_{ij} = \frac{a_{ij}}{|a_i\neq0|}$, where $|\cdot|$ denotes the cardinality of a set. Therefore, $(\bm{WX})_{i.}$ contains the average for each variable over the neighbours of location $i$. 

\subsection{Accounting for asymmetric effects}
\begin{figure}[H]
\centering
\includegraphics[width=0.5\textwidth]{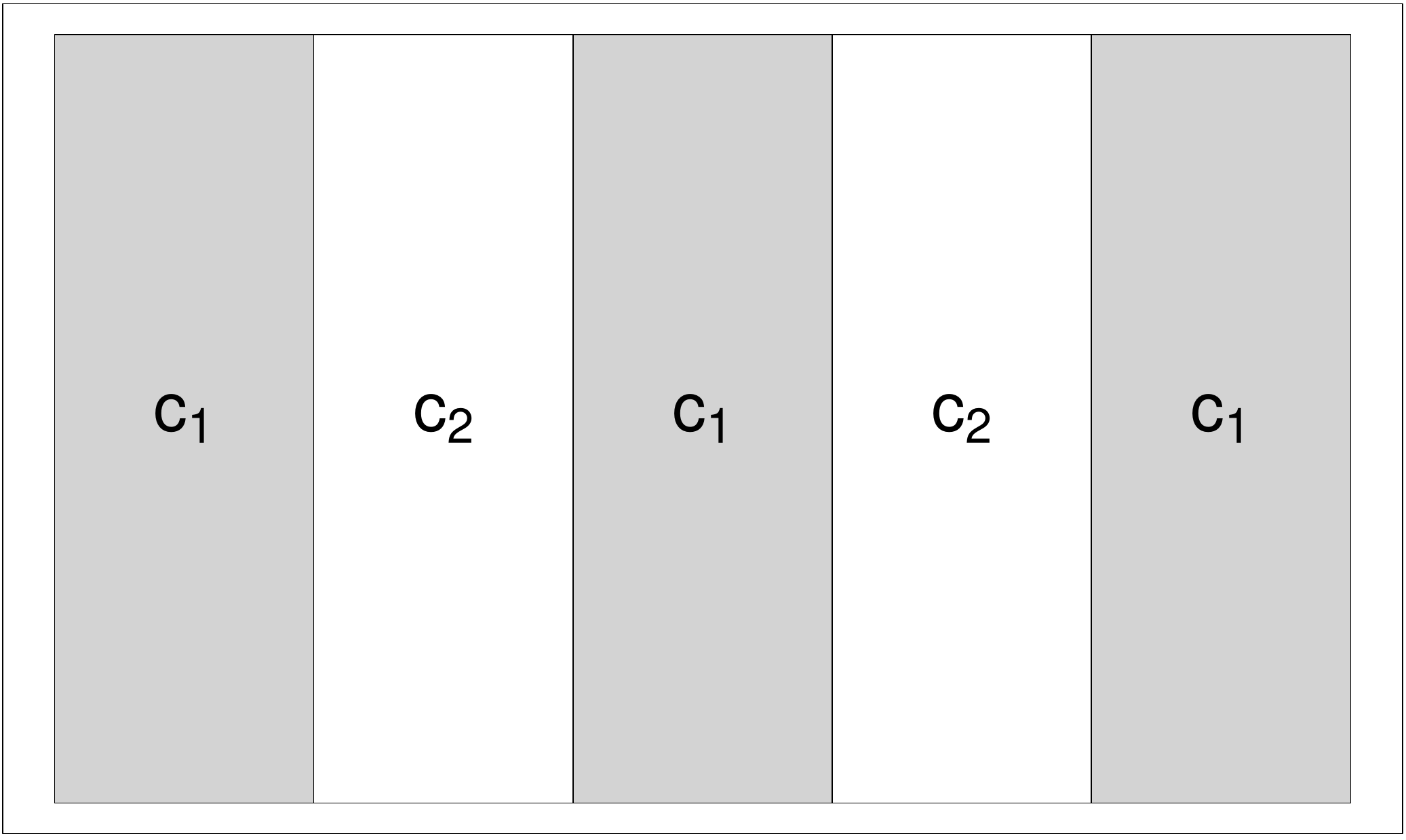}
\caption{Toy example with two different categories, $c_1$ and $c_2$. Each rectangle represents a location.}
\label{fig:intercrop}
\end{figure}

Whenever locations in spatial data can be assigned a category, such that spatial effects between locations pertaining to different categories are expected to be asymmetric, the model proposed in (\ref{eq:msar}) is unsatisfactory. We illustrate this categorisation of locations and the proposed method by means of a toy example consisting of two categories: $c_1$ and $c_2$, which is shown in Figure \ref{fig:intercrop}. In this toy example, we expect asymmetric spatial effects to occur between neighbours with a differing category label. 

\noindent The asymmetry in spatial effects found in Figure \ref{fig:intercrop} is accounted for by setting
\begin{equation}
\label{eq:asymmetriccrop}
    \bm{X} = \bm{W}_{c_2,c_1}\bm{X}\bm{\Psi}_{c_2,c_1} + \bm{W}_{c_1,c_2}\bm{X}\bm{\Psi}_{c_1,c_2} + \bm{E},\\
\end{equation}
where $\bm{\Psi}_{c_2,c_1} \in \mathbb{R}^{p \times p}$ contains the spatial effect of each variable of category $c_2$  on each variable of category $c_1$ and $\bm{\Psi}_{c_1,c_2}$ $\in$ $\mathbb{R}^{p \times p}$ contains the spatial effect of each variable of category $c_1$  on each variable of category $c_2$. Spatial weight matrices $\bm{W}_{c_2,c_1} \in [0,1]^{n \times n}$ and $\bm{W}_{c_1,c_2} \in [0,1]^{n \times n}$ are chosen such that only the matrix elements pertaining to directly neighbouring locations of respectively category $c_1$ and $c_2$ containing respectively category $c_2$ and $c_1$ are nonzero. 

\begin{definition} (Spatial weight matrix structure) \label{defspaweightstruc}
Given two neighbouring locations with categories $c_1$ and $c_2$, the adjacency matrix $\bm{A}_{c_2,c_1}$, isolating the observations of category $c_2$ with neighbour category $c_1$, is of the following form: 
\begin{equation*}
a_{c_2,c_1,i,j}
    \begin{cases}
    1,& \parbox[t]{.6\textwidth}{if location $i$ contains category $c_1$ and location $j$ contains category $c_2$ and locations $i$ and $j$ are neighbours}\\
    0,              & \text{otherwise}.
\end{cases}
\end{equation*}
Subsequently, $\bm{W}_{c_2,c_1}$ contains elements $w_{c_2,c_1,i,j} = \frac{a_{c_2,c_1,i,j}}{|a_{c_2,c_1,i}\neq0|}$. 
\end{definition}

\noindent Thus, following Figure \ref{fig:intercrop}, we obtain the following weight matrices
\begin{equation}
\label{eq:weightmatrices}
    \bm{W}_{c_2,c_1} = \begin{pmatrix}
0 & 1 & 0 & 0 & 0\\
0 & 0 & 0 & 0 & 0\\
0 & 0.5 & 0 & 0.5 & 0\\
0 & 0 & 0 & 0 & 0\\
0 & 0 & 0 & 1 & 0
\end{pmatrix} \quad \bm{W}_{c_1,c_2} = \begin{pmatrix}
0 & 0 & 0 & 0 & 0\\
0.5 & 0 & 0.5 & 0 & 0\\
0 & 0 & 0 & 0 & 0\\
0 & 0 & 0.5 & 0 & 0.5\\
0 & 0 & 0 & 0 & 0
\end{pmatrix}.
\end{equation}

\noindent Using this formulation, whilst it might seem feasible to extend the model to scenarios where more than two categories are present in the data, the model becomes unidentifiable, see Section \ref{Identifiability}. Therefore, we will restrict analyses to scenarios consisting of 2 categories.

To ensure that the parameters can be estimated, spatial autoregressive models typically impose restrictions on the values that the spatial autoregressive parameter can take (c.f.\ Ord, \citeyear{ord1975estimation}; Yang \& Lee, \citeyear{yang2017identification}). Our method is no different in this regard.

\begin{assumption} (Stability condition) \label{assumpposdetd}
To estimate the spatial effects, we require that $\det\left(\bm{I}_{np} -\sum_{k = 1}^{2}\bm{\Psi}_{k}^T \otimes \bm{W}_{k}\right)$ $ > 0$, which is guaranteed whenever $\lambda_{\text{min}}\left(\sum_{k = 1}^{2}\bm{\Psi}_{k}^T \otimes \bm{W}_{k}\right) > -1$ and $\lambda_{\text{max}}\left(\sum_{k = 1}^{2}\bm{\Psi}_{k}^T \otimes \bm{W}_{k}\right) < 1$, where $\lambda_{\text{min}}(\cdot)$ and $\lambda_{\text{max}}(\cdot)$ are the minimum and maximum eigenvalues respectively. 
\end{assumption}

\noindent In addition to the spatial effects contained in the $\bm{\Psi}_k$, $k \in \{1,2\}$, where $k$ is used to denote the category ordering, our interest also lies in the within-location effects, contained in $\bm{\Theta}_E = \bm{\Sigma}_E^{-1}$, the precision matrix of $\bm{E}$. In order to obtain these within-location effects, a so-called spatial filter is required.  Let $\bm{\Psi} = \{\bm{\Psi}_1, \bm{\Psi}_2\}$, the model specification in (\ref{eq:asymmetriccrop}) allows for the construction of such a filter
$\bm{R}(\bm{\Psi})\text{vec}(\bm{X}) = \text{vec}(\bm{E})$, where $\bm{R}(\bm{\Psi}) = \left(\bm{I}_{np} -\sum_{k = 1}^{2}\bm{\Psi}_{k}^T \otimes \bm{W}_{k}\right)$ is the spatial filter matrix, filtering out the spatial effects on the observations as $\text{vec}(\bm{X}) - \left(\sum_{k = 1}^{2}\bm{\Psi}_{k}^T \otimes \bm{W}_{k}\right)\text{vec}(\bm{X}) = \text{vec}(\bm{E})$, resulting in the within-location data, assuming that all spatial dependencies are accurately captured by the $\bm{W}_k$ (Getis, \citeyear{getis1990screening}; Millo, \citeyear{millo2014maximum}). This result is straightforwardly obtained from the data generating process, $\text{vec}(\bm{X}) = \bm{R}(\bm{\Psi})^{-1}\text{vec}(\bm{E})$, with $np \times np$ identity matrix $\bm{I}_{np}$, which in turn is obtained by vectorising (\ref{eq:asymmetriccrop}).

On a related note, as $\bm{X}$ is a linear combination of Gaussian variables, $\bm{X}$ itself is Gaussian with $\mathbb{E}[\text{vec}(\bm{X})] = \bm{0}$ and $\text{Var}[\text{vec}(\bm{X})] = \bm{R}(\bm{\Psi})^{-1}( \bm{\Sigma}_E \otimes \bm{I}_n)\bm{R}(\bm{\Psi})^{-T}$, see Supporting Information. Consequently, $\text{vec}(\bm{X})$ is distributed as $\text{vec}(\bm{X}) \sim N_{np}\left(\bm{0}, \bm{R}(\bm{\Psi})^{-1}( \bm{\Sigma}_E \otimes \bm{I}_n)\bm{R}(\bm{\Psi})^{-T}\right)$, shortened as $\text{vec}(\bm{X}) \sim N_{np}(\bm{0},  \bm{\Sigma}_{X})$. The resulting $ \bm{\Sigma}_{X} =  \bm{\Theta}_{X}^{-1}$ matrix is a $np \times np$ block matrix, where each block contains the variances or covariances within and between categories. $\bm{\Theta}_E$ consists of the original conditional dependences within a location, prior to being affected by any spatial processes. As the spatial effects also distort the conditional dependencies within a location, evaluating the within-location elements of $ \bm{\Theta}_{X}$ is inadequate. In essence, in addition to the $\bm{\Psi}_k$, our interest lies in the true underlying conditional dependencies found in $ \bm{\Theta}_E$, even though we only have data after the occurrence of spatial effects. By separating these effects, the true within-location dependencies can be obtained. As such, there is an equivalence between $\bm{\Theta}_E$, containing the within-location interactions, and the precision matrix corresponding to the time series chain graph (Dahlhaus \& Eichler \citeyear{dahlhaus2003causality}; Abegaz \& Wit, \citeyear{abegaz2013sparse}), which contains the contemporaneous interactions, in addition to establishing the connection to the Gaussian graphical model.

\subsection{Graphical model}
Gaussian graphical models are multivariate statistical models that use graphs $\mathcal{G} = (\mathcal{V},\mathcal{E})$ to represent the full conditional dependence structure between variables represented by a set of vertices $\mathcal{V} = \{1,2,\ldots,p\}$ through the use of a set of undirected edges $\mathcal{E} \subset \mathcal{V} \times \mathcal{V}$, and depends on a $n \times p$ data matrix $\bm{X} = (\bm{X}_1, \bm{X}_2,\ldots,X_p), \bm{X}_j = (X_{1j}, X_{2j}, \ldots, X_{nj})^T, j = 1,\ldots,p$, where the $n$ row vectors in $\bm{X}$ are independent and identically distributed according to $N_p(\bm{0}, \bm{\Theta}^{-1})$. $\bm{\Theta}$ contains the scaled partial correlations, where the partial correlations are given by: $\rho_{ij} = -\frac{\theta_{ij}}{\sqrt{\theta_{ii}\theta_{jj}}}$. Therefore, $(i,j) \not\in \mathcal{E} \Leftrightarrow \theta_{ij} = 0$. 

The spatial autoregressive graphical model is a special type of Gaussian graphical model consisting of both directed and undirected edges. This graphical model captures the complex dependency structure of the data in a parsimonious and well-interpretable manner. To formalise this notion, a few definitions are introduced.

\begin{definition} (Within-location effect graph) \label{defwpe}
The within-location graph corresponding to $\bm{X}$ is an undirected graph $\mathcal{G}_{\w} = (\mathcal{V}, \mathcal{E}_{\w} )$ with edge set $\mathcal{E}_{\w}$ such that
\begin{equation*}
    (i \textrm{---} j)  \not\in \mathcal{E}_{\w} \Leftrightarrow  \theta_{Eij} =  \theta_{Eji} = 0, \quad 1\leq i \neq j \leq p.
\end{equation*}
\end{definition}

\begin{definition} (Between-location effect graph) \label{defne}
The between-location effect graph corresponding to $\bm{X}$ is a directed graph $\mathcal{G}_{\bb:c_1,c_2} = (\mathcal{V}_{c_1} \cup \mathcal{V}_{c_2}, \mathcal{E}_{\bb:c_1,c_2} \cup \mathcal{E}_{\bb:c_2,c_1}) = \mathcal{G}_{\bb:c_2,c_1}$, where $c_2 \in \mathcal{N}_{c_1}$, with $\mathcal{N}_{c_1}$ denoting the set of categories neighboured to category $c_1$, and with edge set $\mathcal{E}_{\bb:c_1,c_2} \cup \mathcal{E}_{\bb:c_2,c_1}$ such that
\begin{equation*}
\begin{gathered}
    (i \rightarrow j)  \not\in \mathcal{E}_{\bb:c_1,c_2} \Leftrightarrow \psi_{c_1,c_2,ij} = 0, \quad
    (j \rightarrow i)  \not\in \mathcal{E}_{\bb:c_1,c_2} \Leftrightarrow \psi_{c_1,c_2,ji} = 0,\\
    (i \rightarrow j)  \not\in \mathcal{E}_{\bb:c_2,c_1} \Leftrightarrow \psi_{c_2,c_1,ij} = 0, \quad
    (j \rightarrow i)  \not\in \mathcal{E}_{\bb:c_2,c_1} \Leftrightarrow \psi_{c_2,c_1,ji} = 0.
\end{gathered}
\end{equation*}
Therefore, $\mathcal{E}_{\bb:c_1,c_2}$ represents the set of all directed edges from category $c_1$ to category $c_2$ and $\mathcal{E}_{\bb:c_2,c_1}$ represents the set of all directed edges from category $c_2$ to category $c_1$. 
\end{definition}

\noindent From Definition \ref{defwpe} we note that $ \theta_{Eij}$ is the scaled partial correlation between variables $i$ and $j$ within any location and Definition \ref{defne} shows that $\psi_{c_1,c_2,ij}$ is the spatial effect of variable $i$ of category $c_1$ on the value of variable $j$ of category $c_2$. Directed edges in the method proposed by Dahlhaus and Eichler (\citeyear{dahlhaus2003causality}) reflect Granger causality. This is, however, not the case for the method that we propose, because almost all spatial orderings are multi-dimensional, and one can move back and forth between locations. 

By combining Definitions \ref{defwpe} and \ref{defne}, a spatial chain graph is obtained that contains the full between- and within-location dependency structure. 

\begin{definition} (Spatial chain graph) \label{defchain}
The spatial chain graph corresponding to $\bm{X}$ is a partially directed graph $\mathcal{G}_{\s} = (\mathcal{V}_{\s}, \mathcal{E}_{\s}) = (\mathcal{V}_{c_1} \cup \mathcal{V}_{c_2}, \mathcal{E}_{\bb:c_1,c_2} \cup \mathcal{E}_{\bb:c_2,c_1} \cup \mathcal{E}_{\w})$ with edge set $\mathcal{E}_{\bb:c_1,c_2} \cup \mathcal{E}_{\bb:c_2,c_1} \cup \mathcal{E}_{\w}$ as defined in Definitions \ref{defwpe} and \ref{defne} respectively. Thus, for $c_2 \in \mathcal{N}_{c_1}$, there exists a directed edge from $\mathcal{V}_{i,c_1}$ to $\mathcal{V}_{j,c_2}$ if and only if $(i \rightarrow j) \in \mathcal{E}_{\bb:c_1,c_2}, 1\leq i \neq j \leq p$. Moreover, for any category $c_1$, there exists an undirected edge between $\mathcal{V}_{i,c_1}$ and $\mathcal{V}_{j,c_1}$ if and only if $(i \textrm{---} j) \in \mathcal{E}_{\w}$.
\end{definition}

\noindent As an example, consider Equation (\ref{eq:asymmetriccrop}) pertaining to the toy example illustrated in Figure \ref{fig:intercrop}. Suppose that $p = 4$ and that we have obtained the following parameter estimates

\begin{equation*}
    \hat{\bm{\Psi}}_{c_1,c_2} = \begin{pmatrix}
\psi_{11} & 0 & 0 & \psi_{14}\\
0 & 0 & \psi_{23} & 0\\
0 & 0 & 0 & \psi_{34}\\
0 & 0 & 0 & \psi_{44}
\end{pmatrix} \quad \hat{\bm{\Psi}}_{c_2,c_1} = \begin{pmatrix}
0 & \psi_{12} & 0 & 0\\
0 & \psi_{22} & 0 & \psi_{24}\\
0 & 0 & 0 & 0\\
0 & 0 & 0 & 0
\end{pmatrix},
\quad  \hat{\bm{\Theta}}_E = \begin{pmatrix}
 \theta_{11} & 0 &  \theta_{13} &  \theta_{14}\\
0 &  \theta_{22} &  \theta_{23} & 0\\
 \theta_{31} &  \theta_{32} &  \theta_{33} &  \theta_{34}\\
 \theta_{41} & 0 &  \theta_{43} &  \theta_{44}
\end{pmatrix}.
\end{equation*}

\noindent Where we dropped the subscripts of the matrix elements for improved readability. 

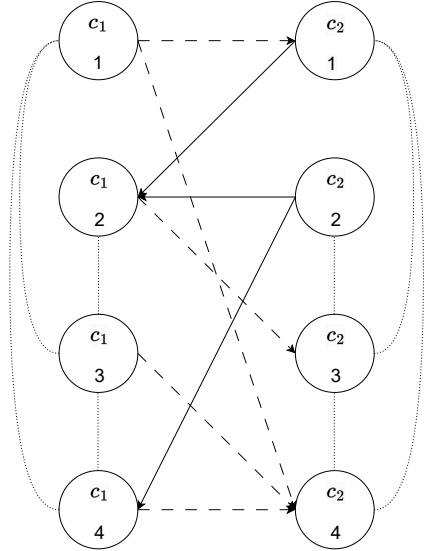
\begin{figure}[H]
\centering
\includegraphics[width=0.33\textwidth]{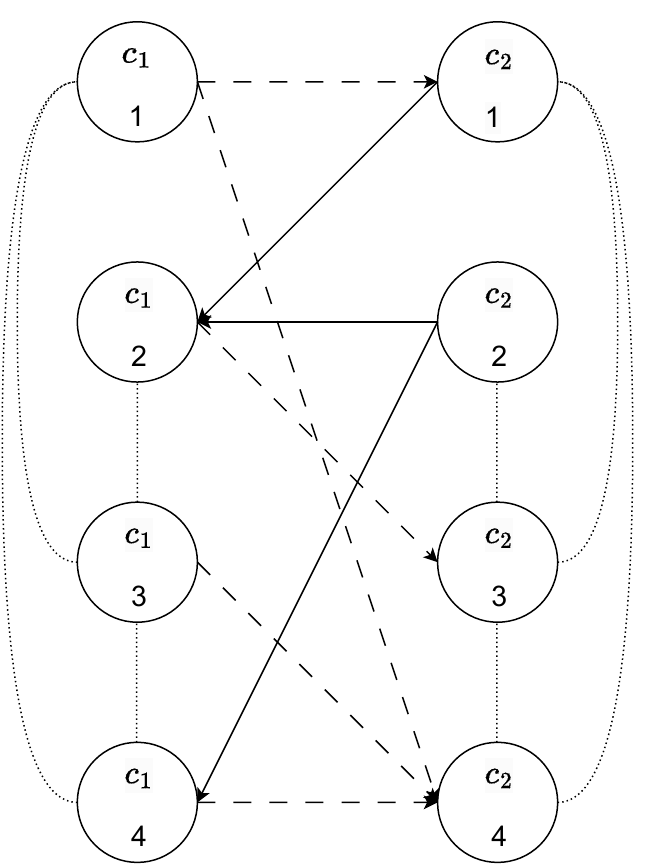}
\caption[]{Spatial chain graph, as derived from $ \bm{\Theta}_E$ (\begin{tikzpicture}\draw [dotted] (0,2) -- (1,2);\end{tikzpicture}), $\bm{\Psi}_{c_1,c_2}$ (\begin{tikzpicture}\draw [->, dashed] (0,2) -- (1,2);\end{tikzpicture}) and $\bm{\Psi}_{c_2,c_1}$ (\begin{tikzpicture}\draw [<-, thick] (0,2) -- (1,2);\end{tikzpicture}) from Equation (\ref{eq:asymmetriccrop}). The graph represents the data generating process underlying observations across two categories, $c_1$ and $c_2$, with 4 variables.}
\label{fig:toy}
\end{figure}

\noindent To illustrate the properties of $ \bm{\Theta}_E$ and the different $\bm{\Psi}_k$, consider Figure \ref{fig:toy}. One of the conditional independence relations that exist within each plot found in the example is the one where variables 1 and 2 are conditionally independent. However, between neighbouring locations, there exists direct spatial effects between these variables. By “direct spatial effects", we refer to any effect from one plots to its first-order neighbours, as will be illustrated in Section \ref{Conditional independence relations}. 

\subsection{Independence relations} \label{Conditional independence relations}
A graphical model for a set of variables is defined as a model where the conditional dependence relations specified by the model are given by a graph. Accordingly, a graphical model goes beyond the visualisation of relationships found in data. This also holds for the  spatial autoregressive graphical model, where a distinction should be made between two types of dependence relations: within- and between-category dependencies. Accordingly, this section will cover both type of dependencies and illustrate how they hold for the proposed method. Proofs are presented in the Supporting Information.

\begin{theorem}(Within-category conditional independence)
\label{withingmp}
Let $\mathcal{G}_{\s}$ be a spatial chain graph according to Definition \ref{defchain} corresponding to $\bm{X}$ with mutually disjoint sets $A_c \cup B_c \cup S_c \subseteq \bm{X}_c \subset \bm{X}$, where $\bm{X}_c$ are the observations pertaining to category $c$. Then, if $S_c$ separates $A_c$ and $B_c$ in the spatial chain graph, we have that 
\begin{equation*}
    A_c \indep B_c|S_c.
\end{equation*}
\end{theorem}

\noindent Given that the global Markov property holds for any category in the spatial chain graph, the local and pairwise Markov properties hold as well (Lauritzen, \citeyear{lauritzen1996graphical}).
\\
\\
In line with the typical spatial autoregressive models that do not exhibit independence between non-first-order neighbours, we show that the proposed method does not either. Let the data generating process be $\text{vec}(\bm{X}) = \left(\bm{I}_{np} - \sum_{k = 1}^{2} \bm{\Psi}_k^T \otimes \bm{W}_k\right)^{-1}\text{vec}(\bm{E})$. The inverse term is what causes dependencies between a location and its non-first-order neighbours through the residuals as $\left(\bm{I}_{np} - \sum_{k = 1}^{2} \bm{\Psi}_k^T \otimes \bm{W}_k\right)^{-1}\text{vec}(\bm{E}) = \left[\bm{I}_{np} + \sum_{q = 1}^{\infty}\left(\sum_{k = 1}^{2}\bm{\Psi}_k^T \otimes \bm{W}_k\right)^q\right]\text{vec}(\bm{E})$, dependencies between a location and its non-first-order neighbours exist. Naturally, the first order term $\left(\sum_{k = 1}^{2}\bm{\Psi}_k^T \otimes \bm{W}_k\right)^1$ results in a non-zero matrix, but so do the higher order terms, causing spillovers to non-neighbouring locations of the residual term $\bm{E}$. However, we have that $\left(\sum_{k = 1}^{2}\bm{\Psi}_k^T \otimes \bm{W}_k\right)^q < \left(\sum_{k = 1}^{2}\bm{\Psi}_k^T \otimes \bm{W}_k\right)^{q-1}$, with $q \in \mathbb{N}^{+}$, due to the restriction imposed by Assumption \ref{assumpposdetd}. This result implies that the strength of the spatial effect decreases as the neighbour order increases. The spillovers occur only within sequences of adjacent locations, not between sequences of adjacent locations. From an applied perspective, the spillovers make sense: whilst most of the interactions between-locations occur between those that are adjacent to one another, it is impossible to guarantee complete isolation from non-adjacent locations. Moreover, the further away locations are, the less strong the spatial effect. However, these spillovers happen only between categories that share at least 1 border on their respective locations. Due to identifiability issues, the proposed method cannot handle more than 2 categories. In spite of this, for data consisting of more than 2 categories, any pair of categories that does not share a border with another pair of categories are independent, implying that the method can even handle spatial “big data”, as long as the combinations are analysed separately. This is shown in the following Theorem. 

\begin{theorem}(Between-category independence: non-neighbours)
\label{betweengmpcond}
Let $\mathcal{G}_{\s}$ and $\mathcal{G}_{\s'}$ be spatial chain graphs according to Definition \ref{defchain} corresponding to $\bm{X}_{c_1,c_2}$ and $\bm{X}_{c_3,c_4}$ respectively, such that $\bm{X}_{c_1,c_2} \cup \bm{X}_{c_3,c_4} \subseteq \bm{X}$, with disjoint sets $A_{c_1,c_2} \subseteq \bm{X}_{c_1,c_2}$ and $B_{c_3,c_4} \subseteq \bm{X}_{c_3,c_4}$, where $c_2 \in \mathcal{N}_{c_1}$ and $c_4 \in \mathcal{N}_{c_3}$. Then, if $c_1,c_2 \notin \mathcal{N}_{c_3} \cup \mathcal{N}_{c_4}$ we have that
\begin{equation*}
    A_{c_1,c_2} \indep B_{c_3,c_4}.
\end{equation*}
\end{theorem}

\noindent The result from this theorem leads to the following special case for neighbouring categories.

\begin{corollary}(Between-category independence: no spatial effect)
\label{betweengmpse}
Let $\mathcal{G}_{\s}$ be a spatial chain graph according to Definition \ref{defchain} corresponding to $\bm{X}$ with disjoint sets $A_{c_1} \subseteq \bm{X}_{c_1}$ and $B_{c_2} \subseteq \bm{X}_{c_2}$, such that $\bm{X}_{c_1} \cup \bm{X}_{c_2} \subseteq \bm{X}$, where $c_1 \neq c_2$. Then, if $c_2 \in \mathcal{N}_{c_1}$ and $\bm{\Psi}_{c_2,c_1} = \bm{\Psi}_{c_1,c_2} = \bm{O}_p$ we have that 
\begin{equation*}
    A_{c_1} \indep B_{c_2}.
\end{equation*}
\end{corollary}

\subsection{Identifiability} \label{Identifiability}
The popular adage “there is no such thing as a free lunch", holds true even in statistics. Given that the proposed model is highly parameterised, identification is a problem. This problem is illustrated by means of an example in the Supporting Information. Due to the non-identifiability of the model, estimation is a non-trivial issue; one that is not often considered in the Bayesian setting, as informative priors have the potential to circumvent identification issues (Gelfand \& Sahu, \citeyear{gelfand1999identifiability}). For the proposed method, we allow for both informative priors and parameter restrictions such that model identification, and in turn Bayesian learning, becomes possible. In the proposed method, we have that the model is identifiable for $\bm{\Theta}_E$, but not for the various $\bm{\Psi}_k$, without informative priors or additional parameter restrictions. To this end, the following parameter restrictions are proposed

\begin{assumption} (Identifiability restrictions) \label{ass:identifiability}
To identify the spatial effects, we impose the following restrictions on the various $\bm{\Psi}_k$ (i) $\psi_{kii} \sim N(\mu,\tau)$, for all $i = 1,\ldots,p$, and for small $\tau > 0$ and $\mu \in \mathbb{R}$ and (ii) $\bm{\Psi}_k = \bm{\Psi}_k^T$ or (iii) the $\bm{\Psi}_k$ are (near) triangular matrices where the 0-elements are assigned the same $N(0,\tau)$ priors as the diagonal. Alternatively, we can impose restriction (iv) where $N(\mu,\tau)$ priors are imposed on known effects of the $\bm{\Psi}_k$ for $1 \leq k \leq 2$. \end{assumption}

\noindent Note that we require both restrictions (i) and (ii) or (i) and (iii) to obtain identifiability, as without these restrictions, or informative priors, regardless of the number of observations, the parameters cannot be estimated. However, restriction (iv) is sufficient for identifiability, provided there exists at least $p^2/2$ known effects per $\bm{\Psi}_k$. As such, “restriction” (iv) is not so much of a restriction, but rather a typical example of Bayesian statistics whereby knowledge is translated into priors, in order to improve statistical inference. However, in this case, inference is not merely improved, but made possible by using informative priors. Conversely, no identifiability restrictions are needed if the same set of spatial effects is assumed for all locations, that is, the flexibility of asymmetric spatial relations is abandoned and, instead, a single $\bm{\Psi}$ is estimated. Whilst the lack of identifiability restrictions appears to be appealing for some spatial processes of which no prior knowledge exists, one of the primary strengths of the proposed method is to model asymmetric spatial effects. As such, this option will not be pursued further in this article, and prior knowledge of some of the spatial effects is assumed. A straightforward example of prior knowledge presents itself in intercropping data consisting of production (e.g.\ yield), environmental and management variables. Whilst management variables might affect production variables, it is a-priori known that the reverse direction contains no effect, resulting in a tight prior around 0 for this value. It is up to the researcher to choose the appropriate restriction, which varies on a case-by-case basis. 

As identification issues remain for data with more than 2 categories, we recommend that the parameters corresponding to each category combination are estimated separately from the other combinations by fitting a different model for each combination. Therefore, we cannot pool the observations to estimate the common precision matrix, but obtain different $\bm{\Theta}_E$ for each combination, whose differences disappear asymptotically. As such, by Theorem \ref{betweengmpcond}, asymptotically, it is irrelevant for the $\bm{\Psi}_k$ whether they are estimated in a joint-fashion, or whether we estimate them separately for each combination. It is impossible to further generalise Theorem 2 to separately estimate the $\bm{\Psi}_k$, thereby loosening the restrictions imposed in Assumption (\ref{ass:identifiability}), as shown in the Supporting Information.

\section{Bayesian inference} \label{Bayesian inference}
A Bayesian framework for the spatial autoregressive model consists of two choices: the prior choice for the autoregressive parameters and the prior choice for the within-location precision matrix. In fact, an additional consideration for the autoregressive parameters needs to be made: whether or not sparsity-inducing priors are imposed. Sparsity --- having a majority of zero coefficients --- has seen a surge of interest over the last few years, including in the spatial autoregressive literature (Pfarrhofer \& Piribauer, \citeyear{pfarrhofer2019flexible}), and has various advantages over priors that result in non-sparse estimates, such as selecting only relevant variables in high-dimensional situations ($n \ll p$) for improved predictive accuracy, increased interpretability by forcing certain coefficients to zero and faster parameter estimation. To ensure that the proposed method is generalisable, we allow for both non-sparse and sparse spatial autoregressive priors. For the former. we choose the normal prior (LeSage \& Chih, \citeyear{lesage2018bayesian}) and for the latter the normal-gamma prior. The normal-gamma prior has shown good performance under conditions of (intermediate) sparsity (Huber \& Feldkircher \citeyear{huber2019adaptive}; Kastner \& Huber, \citeyear{kastner2020sparse}). 

Even though elaborate prior structures are considered for the residual covariance matrix in the aforementioned articles, they are not applicable to the proposed model, as they do not lend themselves to structure learning of the within-location effect graph, as the graph space remains unexplored. To learn the graph structure of a graphical model, researchers typically impose the conjugate G-Wishart prior on the precision matrix and assume a uniform or truncated Poisson on the graph space (Dawid \& Lauritzen, \citeyear{dawid1993hyper}; Roverato, \citeyear{roverato2002hyper}; Mohammadi \& Wit, \citeyear{mohammadi2015bayesian}). However, unless a decomposable graph is assumed, which is atypical, this prior requires computationally intensive evaluation of a normalising constant, which hinders generalisability to big or high-dimensional data. For this reason, Wang (\citeyear{wang2012bayesian}) introduced the Bayesian graphical lasso; an efficient framework to perform Bayesian inference for the graphical lasso. An improvement of the Bayesian graphical lasso has recently been proposed by Li et al.\ (\citeyear{li2019graphical}): the graphical horseshoe prior. This prior has shown excellent performance in graph structure learning, whilst remaining computationally efficient and will be our prior choice on $\bm{\Theta}_E$. Even though the horseshoe prior is known to be conservative in terms of variable selection (van der Pas et al., \citeyear{van2017uncertainty}), i.e.\ few true zero parameters in the model are falsely selected, whilst some of the true non-zero parameters are not selected, Li et al.\ (\citeyear{li2019graphical}) mitigated the number of false negatives by using the 50$\%$ credible interval for variable selection. These priors are implemented in an efficient Gibbs sampling algorithm, that allows researchers to estimate the parameters corresponding to the spatial chain graph for small, moderate or large $p$ in a reasonable amount of time. We will only illustrate the Gibbs sampler for the normal-gamma prior on the autoregressive parameters and not for the normal prior, as its implementation is trivial. Nonetheless, we will illustrate the performance of all priors in Section \ref{Simulation study}. 
\\
\\
The first step of our Bayesian estimation procedure is the likelihood $L(\bm{\Psi}, \bm{\Theta}_E) = \prod_{i = 1}^n P(\bm{x}_i|\bm{\Psi}, \bm{\Theta}_E)$, which under the assumption of multivariate normality of the residuals in (\ref{eq:asymmetriccrop}), is given by 
\begin{gather}
            (2\pi)^{-\frac{np}{2}} \det{(\bm{\Theta}_E)}^{\frac{n}{2}} \det{[\bm{R}(\bm{\Psi})]} \text{exp}\left\{-\frac{1}{2}[\bm{R}(\bm{\Psi})\text{vec}(\bm{X})]^T\left(\bm{\Theta}_E \otimes \bm{I}_n\right)\bm{R}(\bm{\Psi})\text{vec}(\bm{X})\right\}\notag\\
            \propto \det{(\bm{\Theta}_E)}^{\frac{n}{2}} \det{[\bm{R}(\bm{\Psi})]} \text{exp}\left[-\frac{1}{2}\text{tr}\left(\bm{S}\bm{\Theta}_E\right)\right]\label{eq:likbayes},
\end{gather}
with $\bm{S} = \left(\bm{X}- \sum_{k = 1}^{2} \bm{W}_{k}\bm{X}\bm{\Psi}_k\right)^T \left(\bm{X}-\sum_{k = 1}^{2} \bm{W}_{k}\bm{X}\bm{\Psi}_k\right)$. A-priori we assume independence of $\bm{\Theta}_E$ and the $\bm{\Psi}_k$. Due to the formulation of the likelihood in (\ref{eq:likbayes}), the block sampling algorithm of Li et al.\ (\citeyear{li2019graphical}) is straightforwardly applied on the estimation of $\bm{\Theta}_E$ when holding $\bm{{\Psi}}_1$ and $\bm{\Psi}_{2}$ constant. Horseshoe priors are imposed on the upper-triangular elements of the precision matrix (due to the symmetry of $\bm{\Theta}_{E}$), and a uniform prior on the diagonal elements. This results in the following element-wise hierarchical structure 
\begin{equation*}
    \begin{gathered}
        \theta_{Eii} \propto 1,\\
        \theta_{Eij: i < j} \sim N(0,\lambda_{ij}^2 \xi^2),\\
        \lambda_{ij: i < j} \sim C^{+}(0,1),\\
        \xi \sim C^{+}(0,1),\\
    \end{gathered}
\end{equation*}      
where the global $\xi$ and local $\lambda_{ij}$ hyperparameters follow a half-Cauchy $C^{+}(0,1)$ distribution. The posterior on $\bm{\Theta}_E$ then has the following form
\begin{equation*}
       P_{\bm{X}}(\bm{\Theta}_{E}|\bm{\Lambda}, \xi) \propto \det{(\bm{\Theta}_{E})}^{\frac{n}{2}}\text{exp}\left[-\frac{1}{2}\text{tr}\left(\bm{S}\bm{\Theta}_{E}\right)\right]\prod_{i < j}\text{exp}\left(-\frac{\theta_{Eij}^2}{2 \lambda_{ij}^2 \xi^2}\right)\mathbb{I}_{\bm{\Theta}_{E} \in \mathcal{M}_p^+},
\end{equation*}      
where $\mathbb{I}$ is an indicator function, $\bm{\Lambda} = (\lambda_{ij}^2)$ and $\mathcal{M}_p^+$ is the space of $p \times p$  positive definite matrices. Sampling from this posterior is not straightforward, and Li et al.\@ (\citeyear{li2019graphical}) propose a modified version of the efficient block Gibbs sampler introduced in Wang (\citeyear{wang2012bayesian}), whereby one column and row of $\bm{\Theta}_E$ are updated at a time. Details on the algorithm are provided in the Supporting Information.

Continuing with the autoregressive parameters, the normal-gamma global–local shrinkage prior leads to the following hierarchical structure
\begin{equation*}
\begin{gathered}
        \psi_{kij}|\alpha_{kij} \sim N(0,2\omega^{-2}\alpha_{kij}),\\
        \alpha_{kij} \sim \text{G}(\kappa,\kappa),\\
        \omega^2 \sim \text{G}(b_0,b_1),
\end{gathered}
\end{equation*}
for $k \in \{1,2\}$ and $1 \leq i,j \leq p$, with local shrinkage parameter $\alpha_{kij}$, global shrinkage parameter $\omega^2$ and hyperparameters $\kappa, b_0$ and $b_1$ specified by the researcher. Both the local and global shrinkage parameters follow a Gamma distribution. Due to the hierarchical nature of these priors, the conditional posteriors are similar to those of Huber and Feldkircher, (\citeyear{huber2019adaptive}) and Pfarrhofer and Piribauer, (\citeyear{pfarrhofer2019flexible}), as the likelihood is not involved, except for the lowest level (that of the $\psi_{kij}$), for which no known distribution exists in our case given the presence of the $\det[\bm{R}(\bm{\Psi})]$ term. Therefore, we use a Metropolis-Hastings (MH) step in the Gibbs sampler to sample from the full conditionals of $\psi_{kij}$. The conditional posteriors of the shrinkage parameters are 
\begin{equation*}
\begin{gathered}
    P(\alpha_{kij}|\psi_{kij}, \omega) \sim \text{GIG}\left(\kappa - \frac{1}{2},\psi_{kij}^2,\kappa\omega^2\right),\\
    P(\omega^2|\alpha_{111},\ldots,\alpha_{2pp}) \sim \text{G}\left(b_0 + \kappa2p^2,b_1 + \frac{\kappa}{2}\sum_{k=1}^{2}\sum_{i = 1}^p\sum_{j = 1}^p \alpha_{kij}\right),
\end{gathered}
\end{equation*}
such that the local shrinkage parameters follow a generalised inverse Gaussian (GIG) distribution and the global shrinkage parameters follow a Gamma distribution. As stated above, to obtain the $\psi_{kij}$, we use a MH step within the Gibbs algorithm, where a new value $\psi_{kij}^{'}$ is proposed $\psi_{kij}^{'} \sim N(\psi_{kij}^{(t-1)}, \zeta)$, for $1 \leq t \leq$ number of iterations, with $\zeta$ being a tuning parameter to increase efficiency of the MH step. To avoid numerical underflow, we take the natural logarithm  of the acceptance rate: if $\log\left[P\left(\bm{X}|\psi_{1ij}^{'},\bm{\Theta}_E^{(t)}, \bm{\Psi}_{1-ij}^{(t-1)},\bm{\Psi}_{2}^{(t-1)}\right)\right] + \log\left[P\left(\psi_{1ij}^{'}\right)\right] >   \log\left[P\left(\bm{X}|\psi_{1ij}^{(t-1)}, \bm{\Theta}_E^{(t)}, \bm{\Psi}_{1-ij}^{(t-1)},\bm{\Psi}_{2}^{(t-1)}\right)\right] + \log\left[P\left(\psi_{1ij}^{(t-1)}\right)\right]$ set $\psi_{kij}^{(t)} = \psi_{kij}^{'}$ and set $\psi_{kij}^{(t)} = \psi_{kij}^{(t-1)}$ otherwise. The hyperparameters in the acceptance step are omitted and we set $k = 1$ to simplify the notation. Note that due to the separable nature of the priors on the individual spatial effect components, combined with the restrictions imposed by Assumption \ref{ass:identifiability}, we end up with respectively $p(p+1)$, $2p^2$ and $2p^2$ MH steps (parameters) for identifiability restrictions (ii), (iii) and (iv) within each iteration of the Gibbs sampling procedure. To guarantee the stability condition imposed on the spatial effects, see Assumption \ref{assumpposdetd}, we redraw values for $\psi_{kij}^{'}$ until the condition is met. The variable selection for the proposed method follows that of Li et al.\ (\citeyear{li2019graphical}), where, if the 50\% posterior credible interval for any element does not contain zero, that element is selected.
 
\section{Simulation study} \label{Simulation study}
To illustrate the performance of the proposed methodology on data of different dimensionality and network structure, a simulation study was conducted. Whilst, to the best of our knowledge, alternative models do not exist, the relevance of this study is found in its capacity to shed light on the adequacy of Bayesian learning of the true parameters underlying the proposed method. 

The three network structures that underlie $\bm{\Theta}_E$ for the simulation study are random, scale-free and star. Whilst our main interest is illustrating to what extend the Gibbs method can estimate the $\bm{\Psi}_k$, multiple networks underlying $\bm{\Theta}_E$ were chosen to represent various types of within-location dependence relations, thereby illustrating the versatility of the proposed method. 

The data is simulated by first constructing $\bm{\Theta}_E$ according to the chosen network type, with an edge probability of 0.2 in the case of the random network, followed by independently drawing the residual vectors $\bm{E}_i \sim N_p(\bm{0}, \bm{\Theta}^{-1}_E)$ for $i = 1,\ldots,n$. Under assumptions of sparsity, i.e.\ the simulations where normal-gamma priors are used, for each $k \in \{1,2\}$ we draw $\lfloor3p(p-1)/8\rfloor$ elements from a $U(-1,1)$ distribution, and randomly assign these to spatial effects $\psi_{kij}$, whilst setting all other $\lceil p(p-1)/8\rceil$ elements to zero. These elements are taken from the triangular $\bm{\Psi}_k$, which are made symmetric afterwards, if the restriction requires this. When normal priors are used, we draw spatial effects $\psi_{kij} \sim U(-1,1)$ for all $k,i,j$. For both priors, the spatial effects are subsequently transformed to adhere to restriction (i) by drawing $\psi_{kii} \sim N(0,\tau)$ as well as either restriction (ii) or restriction (iii) from Assumption \ref{ass:identifiability}. Finally, we set $\text{vec}(\bm{X}) = (\bm{I}_{np} - \sum_{k = 1}^2\bm{\Psi}^T_k \otimes \bm{W}_k)^{-1}\text{vec}(\bm{E})$, where the $\bm{W}_k$ are equivalent to those found in (\ref{eq:weightmatrices}), when generalised to larger values of $n$. 

For each network structure, 20 spatial autoregressive graphical models are fitted for data consisting of 2 categories, with $n \in \{25,50,100\}$ and $p \in \{4,8, 20\}$. We showcase model performance using restrictions (i), (ii) and (iii) of Assumption \ref{ass:identifiability}, where we set all $\bm{\Psi}_k$ to be upper triangular for models adhering to restriction (iii). We do not evaluate the performance of the method with restriction (iv), as it is a generalisation of restriction (iii). The dimensionality of simulated data is consistent with the data commonly found in intercropping designs ($p = 4$ and $p = 8)$. Nevertheless, the proposed method can handle high(er)-dimensional data, as shown by the simulations with $p = 20$. For each parameter combination, we use the first 1000 iterations as burn-in from a total of 2000 iterations. Inference is based on the 1000 posterior draws for all parameters, where we use the posterior mean as a point estimate to compute the discrepancy measures. 

The discrepancy between the true and the estimated parameters are measured using the Frobenius norm: $||\bm{\Theta}_{E}^0 - \hat{\bm{\Theta}}_E||_F$ for $\bm{\Theta}_E$, and the $F_1$ score: $2\frac{\text{precision}\cdot \text{recall}}{\text{precision}+\text{recall}}$ and the root mean square error: 

\noindent $\sqrt{\frac{1}{2p^2}\sum_{k = 1}^{2}\sum_{i = 1}^p\sum_{j = 1}^p (\psi_{kij}^0 - \hat{\psi}_{kij})^2}$ for the $\bm{\Psi}_k$, where the discrepancy measures are averaged over the 20 fitted models. For both the Frobenius norm in $[0,\infty)$ and the root mean square error in $[0,\infty)$, lower values are indicative of better model performance, whilst the opposite holds true for the $F_1$ score in $[0,1]$. The $F_1$ score is computed using the same edge selection method found in Li et al.\ (\citeyear{li2019graphical}). That is, if the 50\% posterior credible interval for any element in the $\bm{\Psi}_k$ does not contain zero, that element is considered a discovery, and vice versa. The results for the simulation studies can be found in Table \ref{tab:simstudy} where normal priors are implemented with hyperparameters $\mu = 0, \sigma = 1$ and normal-gamma priors are implemented with hyperparameters $b_0 = b_1 = 0.01$ and $\kappa = 0.1$ (Huber \& Feldkircher, \citeyear{huber2019adaptive}; Pfarrhofer \& Piribauer, \citeyear{pfarrhofer2019flexible}). For parameter restrictions (i) and (iii), we fix $\tau = 0.001$, reflecting our oracle knowledge on the (by assumption) known effects. Convergence diagnostics of the simulations are shown in the Supporting Information. In addition, the Supporting Information contain an evaluation of the computation time of the proposed method.

\begin{table}[H]
\resizebox{0.92\textwidth}{!}{%
\begin{adjustbox}{angle=270}
\centering
  \begin{threeparttable}
  \caption{Simulation results for the random, scale-free and star networks, using both priors for both symmetric and triangular parameter restrictions. FN stands for Frobenius norm, $F_1$ stands for $F_1$ score and RMSE stands for root mean square error. The discrepancy measures are averaged across 20 fitted models for each parameter combination and rounded to 2 decimals. Standard errors are provided between parentheses.}
  \label{tab:simstudy}
     \begin{tabular}{l| cc | cc | ccc | ccc}
        \toprule
        \midrule
        \multicolumn{1}{c}{}  & \multicolumn{4}{|c|}{Normal prior on $\bm{\Psi}$}  & \multicolumn{6}{c}{Normal-gamma prior on $\bm{\Psi}$}\\
        \midrule
        \multicolumn{1}{c|}{Random network} & \multicolumn{2}{c|}{Symmetric restriction} & \multicolumn{2}{c|}{Triangular restriction}  & \multicolumn{3}{c|}{Symmetric restriction} & \multicolumn{3}{c}{Triangular restriction}\\
        \midrule
\textbf{$n, p$} & \textbf{FN} & \textbf{RMSE} & \textbf{FN} & \textbf{RMSE} & \textbf{FN} & $\textbf{F}_{\bm{1}}$ & \textbf{RMSE} & \textbf{FN} & $\textbf{F}_{\bm{1}}$ & \textbf{RMSE}\\ \midrule
$25, 4$ & 2.60 (0.62) & 0.28 (0.03) & 2.51 (0.57) & 0.22 (0.02) & 2.06 (0.38) & 0.74 (0.03) & 0.36 (0.04) & 2.32 (0.53) & 0.78 (0.01) & 0.25 (0.02)\\
$50, 4$ & 0.89 (0.12) & 0.18 (0.02) & 0.75 (0.06) & 0.15 (0.01)  & 0.91 (0.16) & 0.77 (0.02) & 0.25 (0.03) & 0.97 (0.16) & 0.79 (0.01) & 0.16 (0.01)\\ 
$100, 4$ & 0.89 (0.17) & 0.15 (0.02) & 0.70 (0.10)& 0.11 (0.01)  & 1.14 (0.57) & 0.78 (0.01) & 0.18 (0.02) & 0.66 (0.12) & 0.80 (0.01) & 0.12 (0.01)\\ 
$25, 8$ & 9.03 (1.21) & 0.31 (0.02) & 11.17 (3.80) & 0.27 (0.01) & 20.88 (7.23) & 0.79 (0.01) & 0.37 (0.02) & 11.99 (3.29) & 0.82 (0.01) & 0.35 (0.02)\\ 
$50, 8$ & 6.34 (1.91) & 0.23 (0.02) & 3.35 (1.00) & 0.19 (0.01) & 12.42 (5.70) & 0.81 (0.01) & 0.25 (0.02) & 4.12 (0.89) & 0.83 (0.01) & 0.22 (0.01)\\ 
$100, 8$ & 4.10 (1.12) & 0.21 (0.02) & 2.29 (1.01) & 0.13 (0.00) & 8.86 (5.24) & 0.83 (0.01) & 0.14 (0.01) & 1.86 (0.55) & 0.84 (0.00) & 0.13 (0.01)\\ 
$25, 20$ & 659.08 (228.26) & 0.40 (0.01) & 1081.98 (124.89) & 0.43 (0.01) & 179.81 (63.65) & 0.80 (0.00) & 0.48 (0.02) & 214.65 (78.69) & 0.84 (0.00) & 0.49 (0.02)\\ 
$50, 20$ & 265.35 (103.37) & 0.31 (0.01) & 747.85 (86.70) & 0.36 (0.01) & 89.12 (41.37) & 0.81 (0.01) & 0.35 (0.01) & 154.50 (52.56) & 0.85 (0.00) & 0.37 (0.02)\\
$100, 20$ & 115.76 (68.14) & 0.240 (0.01) & 356.32 (41.19) & 0.26 (0.01) & 36.55 (18.79) & 0.83 (0.00) & 0.23 (0.01) & 64.72 (21.43) & 0.86 (0.00) & 0.23 (0.01)\\
\midrule
        \multicolumn{1}{c|}{Scale-free network}\\
        \midrule
$25, 4$ & 2.38 (0.34) & 0.28 (0.03) & 2.21 (0.39) & 0.23 (0.01) & 3.33 (0.76) & 0.77 (0.02) & 0.35 (0.03) & 4.05 (1.39) & 0.77 (0.02) & 0.23 (0.01)\\ 
$50, 4$ & 1.26 (0.21) & 0.23 (0.02) & 1.09 (0.13) & 0.16 (0.01) & 1.69 (0.46) & 0.78 (0.02) & 0.23 (0.03) & 1.70 (0.42) & 0.79 (0.01) & 0.19 (0.01)\\ 
$100, 4$ & 1.06 (0.20) & 0.15 (0.02) & 0.74 (0.10) & 0.12 (0.01) & 1.66 (0.84) & 0.79 (0.02) & 0.16 (0.02) & 0.95 (0.12) & 0.78 (0.01) & 0.12 (0.01)\\ 
$25, 8$ & 10.89 (1.85) & 0.31 (0.02) & 12.34 (2.84) & 0.29 (0.01) & 10.37 (2.04) & 0.80 (0.01) & 0.40 (0.03) & 11.47 (3.17) & 0.82 (0.01) & 0.38 (0.02)\\ 
$50, 8$ & 6.02 (1.45) & 0.25 (0.02) & 5.03 (1.50) & 0.19 (0.01) & 6.40 (1.79) & 0.81 (0.01) & 0.25 (0.02) & 4.96 (0.93) & 0.84 (0.00) & 0.23 (0.01)\\ 
$100, 8$ & 4.90 (1.97) & 0.19 (0.02) & 1.61 (0.32) & 0.13 (0.00) & 3.66 (0.83) & 0.83 (0.01) & 0.18 (0.02) & 2.11 (0.47) & 0.85 (0.00) & 0.15 (0.01)\\ 
$25, 20$ & 1081.72 (446.70) & 0.4 (0.01) & 1390.24 (188.14) & 0.46 (0.01) & 127.72 (24.95) & 0.79 (0.00) & 0.49 (0.02) & 113.56 (0.00) & 0.83 (0.01) & 0.45 (0.02)\\ 
$50, 20$ & 682.91 (311.92) & 0.34 (0.01) & 733.92 (106.53) & 0.35 (0.01) & 74.38 (12.09) & 0.81 (0.00) & 0.40 (0.02) & 69.45 (11.96) & 0.84 (0.00) & 0.31 (0.01)\\
$100, 20$ & 282.84 (186.90) & 0.23 (0.01) & 412.56 (46.80) & 0.28 (0.01) & 51.85 (8.17) & 0.82 (0.00) & 0.26 (0.01) & 43.90 (7.64) & 0.85 (0.00) & 0.22 (0.01)\\
\midrule
        \multicolumn{1}{c|}{Star network}\\        
        \midrule
$25, 4$ & 2.36 (0.23) & 0.30 (0.03) & 2.98 (0.67) & 0.22 (0.01) & 3.67 (0.79) & 0.74 (0.02) & 0.31 (0.04) & 3.05 (0.58) & 0.78 (0.02) & 0.26 (0.02)\\ 
$50, 4$ & 1.24 (0.20) & 0.20 (0.03) & 1.28 (0.24) & 0.15 (0.01) & 1.94 (0.58) & 0.78 (0.01) & 0.20 (0.02) & 1.39 (0.25) & 0.79 (0.01) & 0.18 (0.01)\\ 
$100, 4$ & 0.95 (0.11) & 0.15 (0.01) & 0.98 (0.14) & 0.11 (0.01) & 1.68 (0.85) & 0.78 (0.02) & 0.17 (0.02) & 0.85 (0.09) & 0.80 (0.01) & 0.11 (0.01)\\ 
$25, 8$ & 12.01 (3.26) & 0.31 (0.02) & 13.52 (3.64) & 0.29 (0.01) & 14.22 (3.49) & 0.79 (0.01) & 0.43 (0.03) & 15.80 (4.64) & 0.83 (0.01) & 0.34 (0.02)\\ 
$50, 8$ & 5.70 (1.44) & 0.23 (0.01) & 4.95 (1.45) & 0.19 (0.01) & 7.06 (2.11) & 0.82 (0.01) & 0.25 (0.02) & 9.57 (5.04) & 0.84 (0.00) & 0.21 (0.01)\\ 
$100, 8$ & 3.39 (0.66) & 0.19 (0.02) & 2.70 (0.76) & 0.13 (0.00) & 6.73 (2.48) & 0.83 (0.01) & 0.20 (0.02) & 5.86 (2.41) & 0.85 (0.00) & 0.17 (0.01)\\ 
$25, 20$ & 596.11 (250.65) & 0.40 (0.01) & 1729.98 (260.41) & 0.44 (0.01) & 127.62 (30.98) & 0.80 (0.00) & 0.49 (0.02) & 216.45 (50.60) & 0.85 (0.00) & 0.51 (0.02)\\ 
$50, 20$ & 243.76 (113.58) & 0.27 (0.02) & 621.94 (174.77) & 0.32 (0.01) & 85.40 (21.56) & 0.81 (0.00) & 0.37 (0.01) & 113.69 (27.83) & 0.86 (0.01) & 0.35 (0.01)\\
$100, 20$ & 144.80 (55.62) & 0.23 (0.01) & 299.62 (84.37) & 0.25 (0.01) & 40.13 (10.12) & 0.82 (0.01) & 0.21 (0.01) & 67.49 (18.26) & 0.87 (0.01) & 0.18 (0.01)\\
\midrule
\bottomrule
     \end{tabular}
  \end{threeparttable}
    \end{adjustbox}}
\end{table}

As observed in Table \ref{tab:simstudy}, for all network types,  prior assumptions, identifiability restrictions and number of parameters, the discrepancy measures for posterior mean point estimates reduce as the sample size of the data grows. For low $p$, the model is able to provide accurate point estimates. However, as $p$ increases, even for higher values of $n$, the discrepancy between the point estimate and the true value is still substantial. However, due to the high number of parameters that need to be estimated, this is no surprise. Whilst the estimation accuracy of the method appears invariant to the underlying network structure, it is not invariant to the identifiability restriction. This result is not particularly remarkable, considering that under the triangular restriction, the lower triangular estimates of the $\bm{\Psi}_k$ contribute next to nothing to the discrepancy measures, whilst under a symmetric restriction these elements contribute equally to the discrepancy measures as the upper triangular elements do. Finally, both priors show satisfactory and similar performance. 

\section{An application on an intercropping system} \label{Real world data example}
\begin{figure}[H]
\centering
\includegraphics[width=0.24\textwidth]{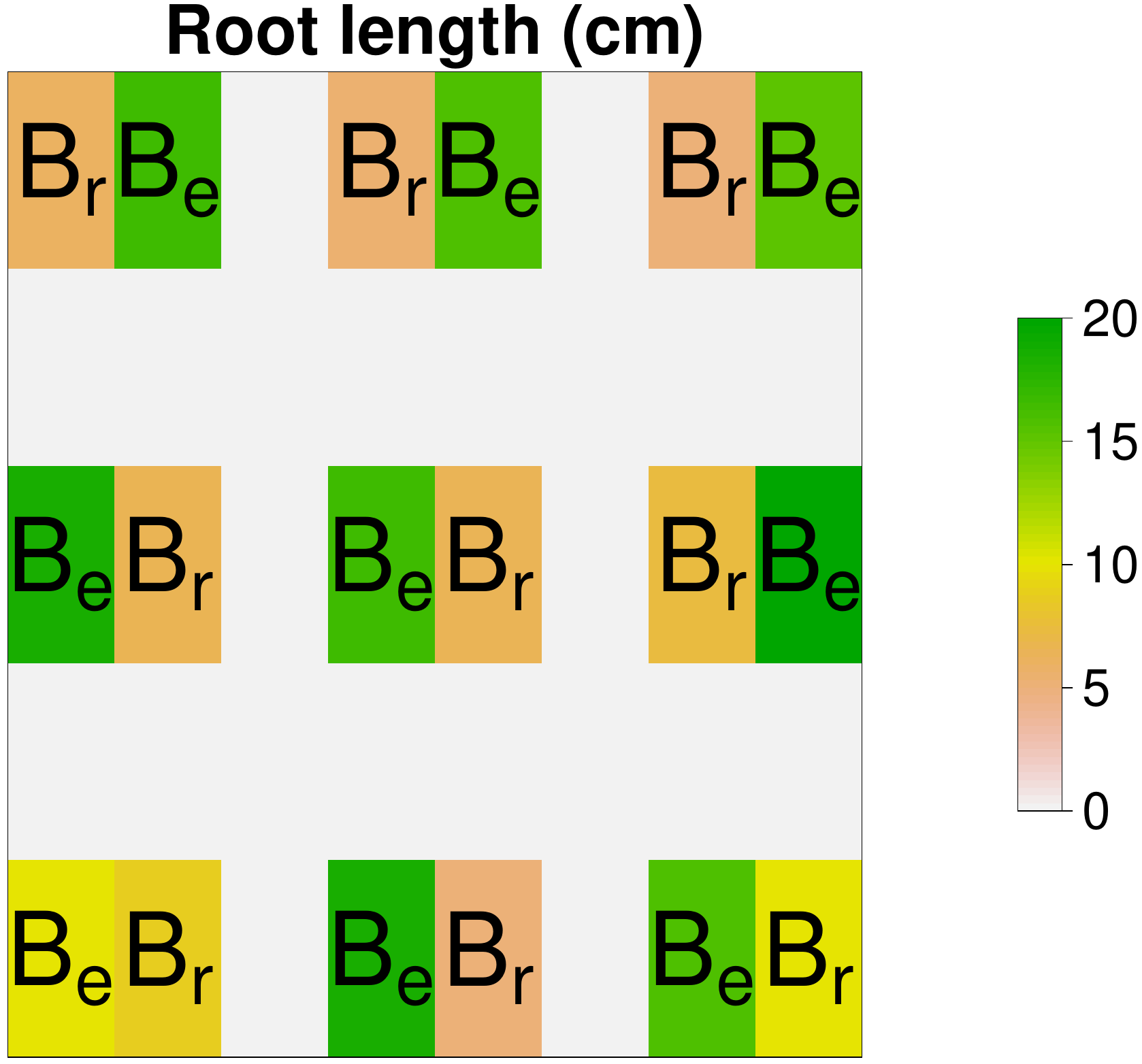}
\includegraphics[width=0.24\textwidth]{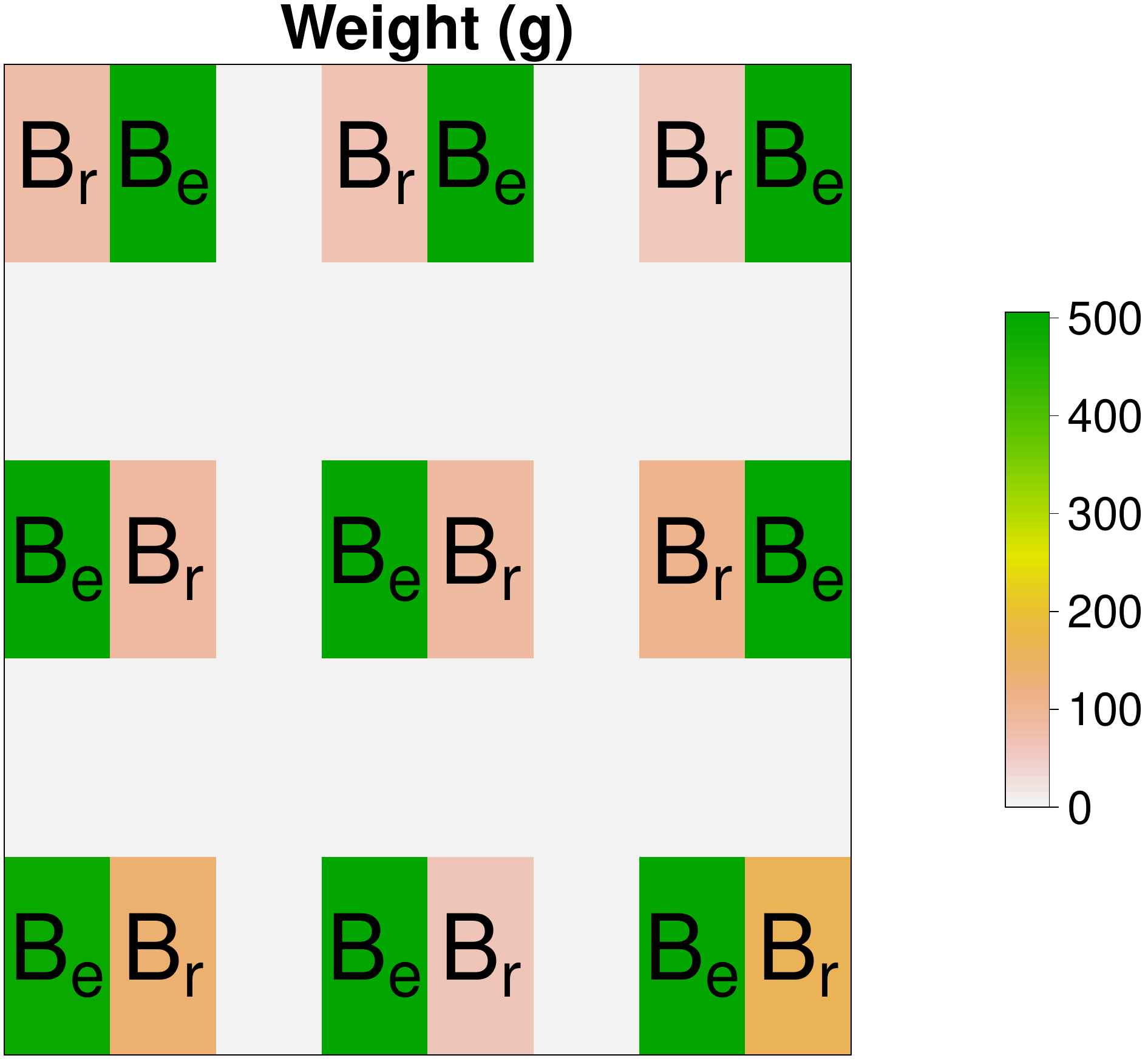}
\includegraphics[width=0.24\textwidth]{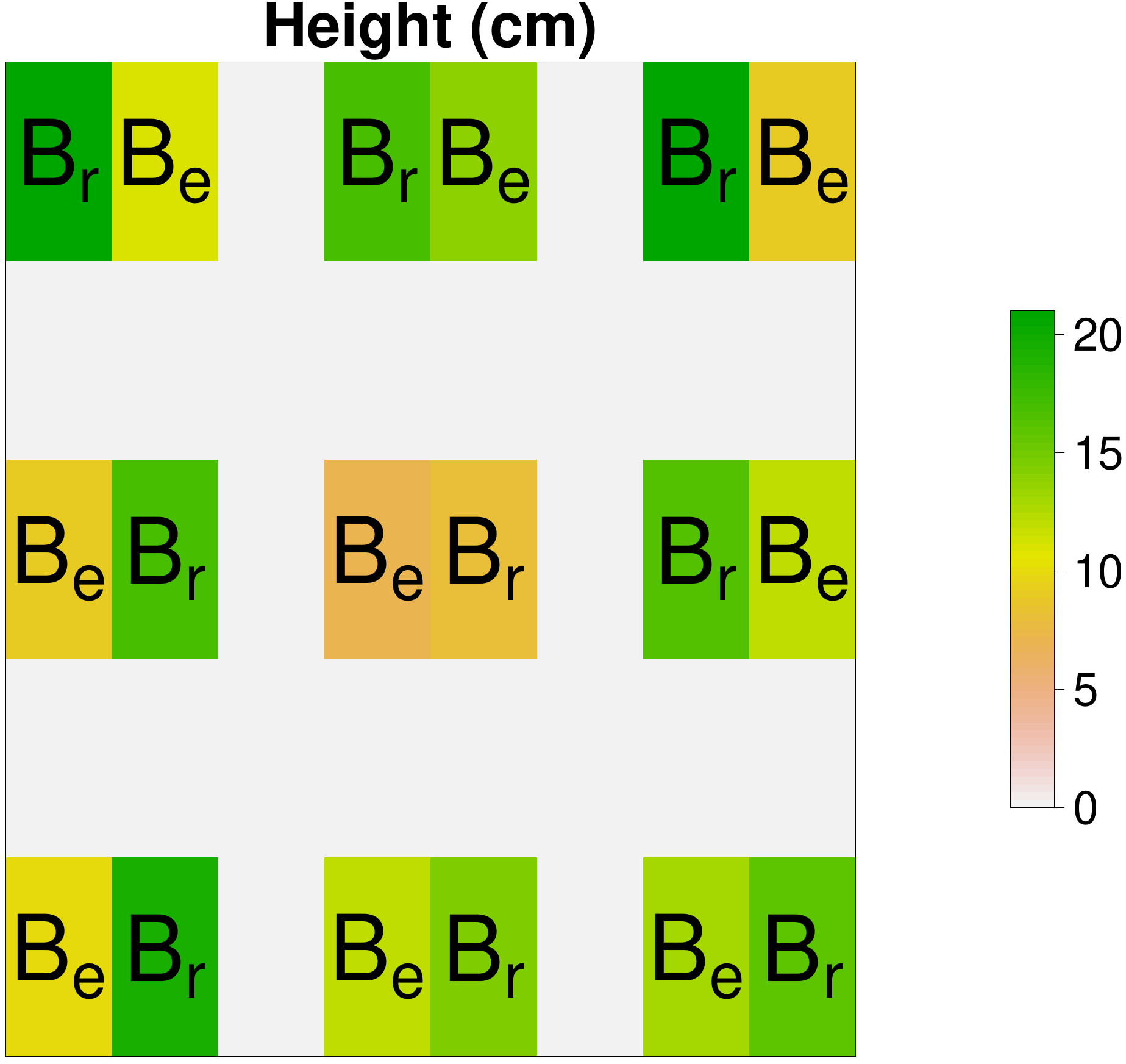}
\includegraphics[width=0.24\textwidth]{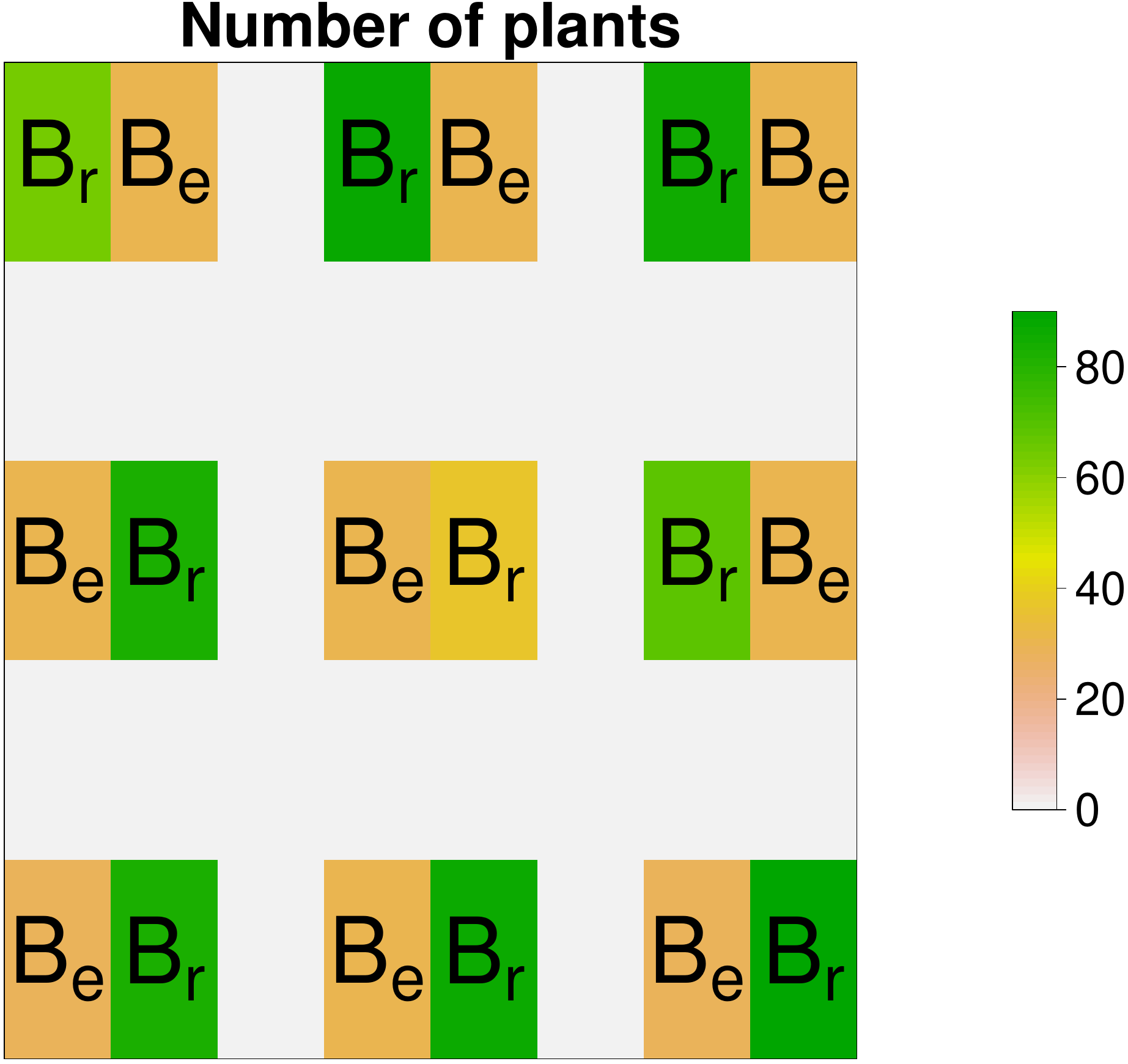}
\caption{Paired crop intercropping design with two different crops: Belgian endive (B\textsubscript{e}) and beetroot (B\textsubscript{r}). Each (coloured) rectangle represents a plot containing multiple plants of the same crop. Variable values for the root length, weight, height and number of plants are illustrated by colours.}
\label{fig:variablevalues}
\end{figure}

We have primarily motivated the present model through an intercropping framework, begging a demonstration on actual intercropping data. The included example consists of a paired-crop type design of Belgian endive and beetroot, where each plot has a single neighbouring plot of the other species. The categories in the proposed method are therefore given by the two different crops. Belgian endive can be found to the right of beetroot 4 times and 5 times to the left, resulting in a total of 18 observations. The design is shown in Figure \ref{fig:variablevalues}, which also contains the variable values for the plots. The variables in this dataset are dried plant weight, fresh plant weight, (aboveground) plant height and number of plants, where the first three variables are averaged across all plants within a plot. Instead of including two different weight variables, we use the fresh plant weight to construct the plant root length based on the average ratio of root weight/total weight for the crops. The different locations represent plots containing multiple plants of a single crop. Given the presence of borders between each pair of neighbouring plots, no spatial spillovers are expected to occur between non-neighbouring plots. Therefore, each pair of Belgian endive and beetroot is independent of all other pairs. The data used in this study was generated through the strategic investment theme ‘Biodiversity-positive Food Systems’ of Wageningen University \& Research (\citeyear{WUR}).

\noindent Whilst the plots containing Belgian endive typically contain more biomass in terms of weight than those containing beetroot, the number of plants per plot, as well as the average plant height tend to be higher for beetroot. 

We use the following equation to represent the stochastic model underlying the observed data
\begin{equation*}
    \bm{X} = \bm{W}_{B_r,B_e}\bm{X}\bm{\Psi}_{B_r,B_e} + \bm{W}_{B_e,B_r}\bm{X}\bm{\Psi}_{B_e,B_r} + \bm{E}, \quad \bm{E}_i \sim N_p(\bm{0},\bm{\Theta}^{-1}_E), \quad i = 1,\ldots,18,
\end{equation*}
where $\bm{W}_{B_r,B_e}, \bm{W}_{B_e,B_r} \in [0,1]^{18 \times 18}$ are of the following form 

\begin{equation*}
    \bm{W}_{B_r,B_e} = \begin{pmatrix}
0 & 0 & 0 & 0 & \dots & 0 & 0\\
1 & 0 & 0 & 0 & \dots& 0 & 0\\
0 & 0 & 0 & 0 & & \vdots & \vdots \\
0 & 0 & 1 & 0 &\ddots \\
\vdots &\vdots & & \ddots& \ddots\\
0 & 0 & \dots & & & 0 & 0\\
0 & 0 & \dots & & & 1 & 0
\end{pmatrix} \quad \bm{W}_{B_e,B_r} = \begin{pmatrix}
0 & 1 & 0 & 0 & \dots & 0 & 0\\
0 & 0 & 0 & 0 & \dots& 0 & 0\\
0 & 0 & 0 & 1 & &  \vdots & \vdots \\
0 & 0 & 0 & 0 &\ddots \\
\vdots &\vdots & & \ddots& \ddots\\
0 & 0 & \dots & & & 0 & 1\\
0 & 0 & \dots & & & 0 & 0
\end{pmatrix}.
\end{equation*}

\begin{figure}[H]
\centering
\includegraphics[width=0.5\textwidth]{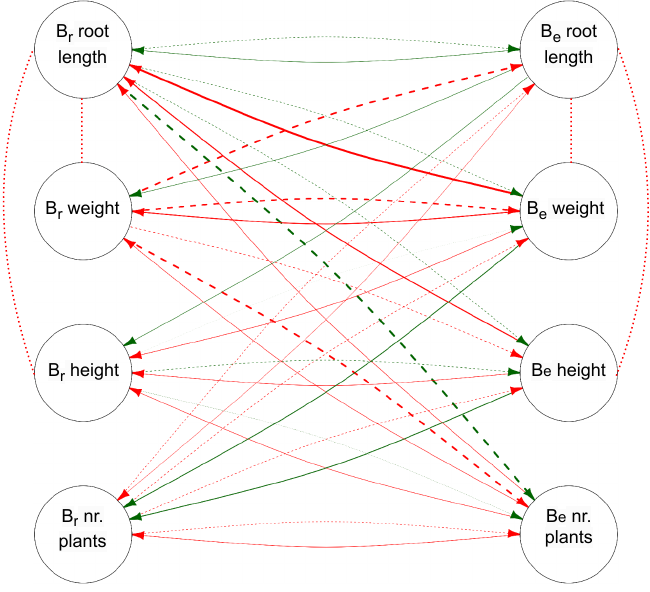}
\caption[]{Spatial chain graph, based on the posterior means from $ \bm{\Theta}_E$ (\begin{tikzpicture}\draw [dotted] (0,2) -- (1,2);\end{tikzpicture}), $\bm{\Psi}_{B_r,B_e}$ (\begin{tikzpicture}\draw [->, dashed] (0,2) -- (1,2);\end{tikzpicture}) and $\bm{\Psi}_{B_e,B_r}$ (\begin{tikzpicture}\draw [<-, thick] (0,2) -- (1,2);\end{tikzpicture}), with positive (\begin{tikzpicture}\draw [colorpos, thick] (0,2) -- (1,2);\end{tikzpicture}) and negative effects (\begin{tikzpicture}\draw [colorneg, thick] (0,2) -- (1,2);\end{tikzpicture}), where the edge width is indicative of the effect size. The graph represents the data generating process underlying observations across two different crops, beetroot (B\textsubscript{r}) and Belgian endive (B\textsubscript{e}), with 4 variables.}
\label{fig:toepassing}
\end{figure}
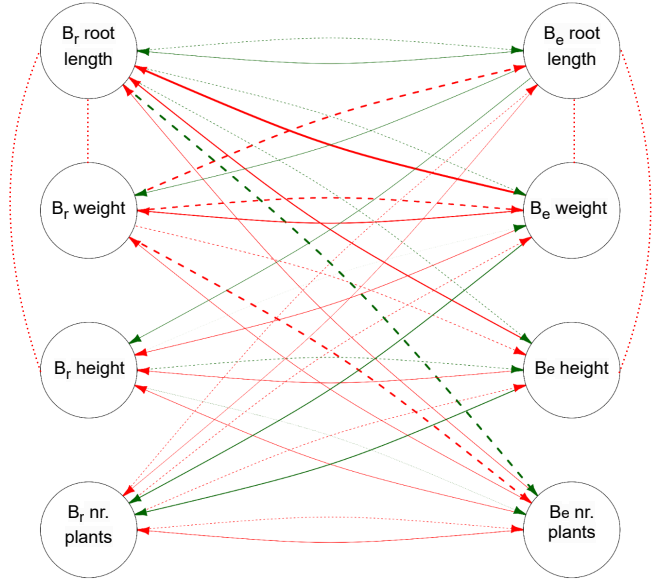

\begin{figure}[H]
\centering
\includegraphics[width=0.24\textwidth]{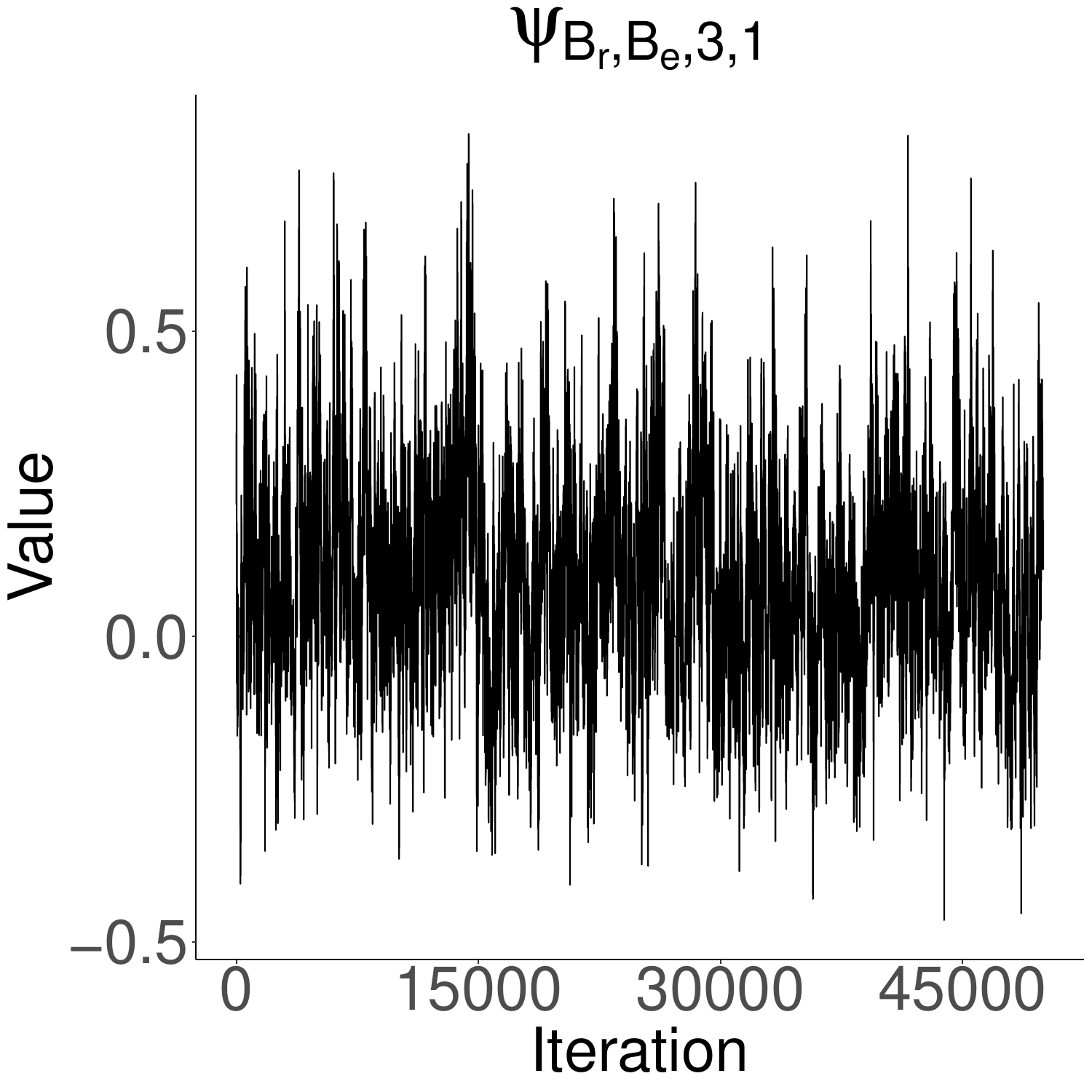}
\includegraphics[width=0.24\textwidth]{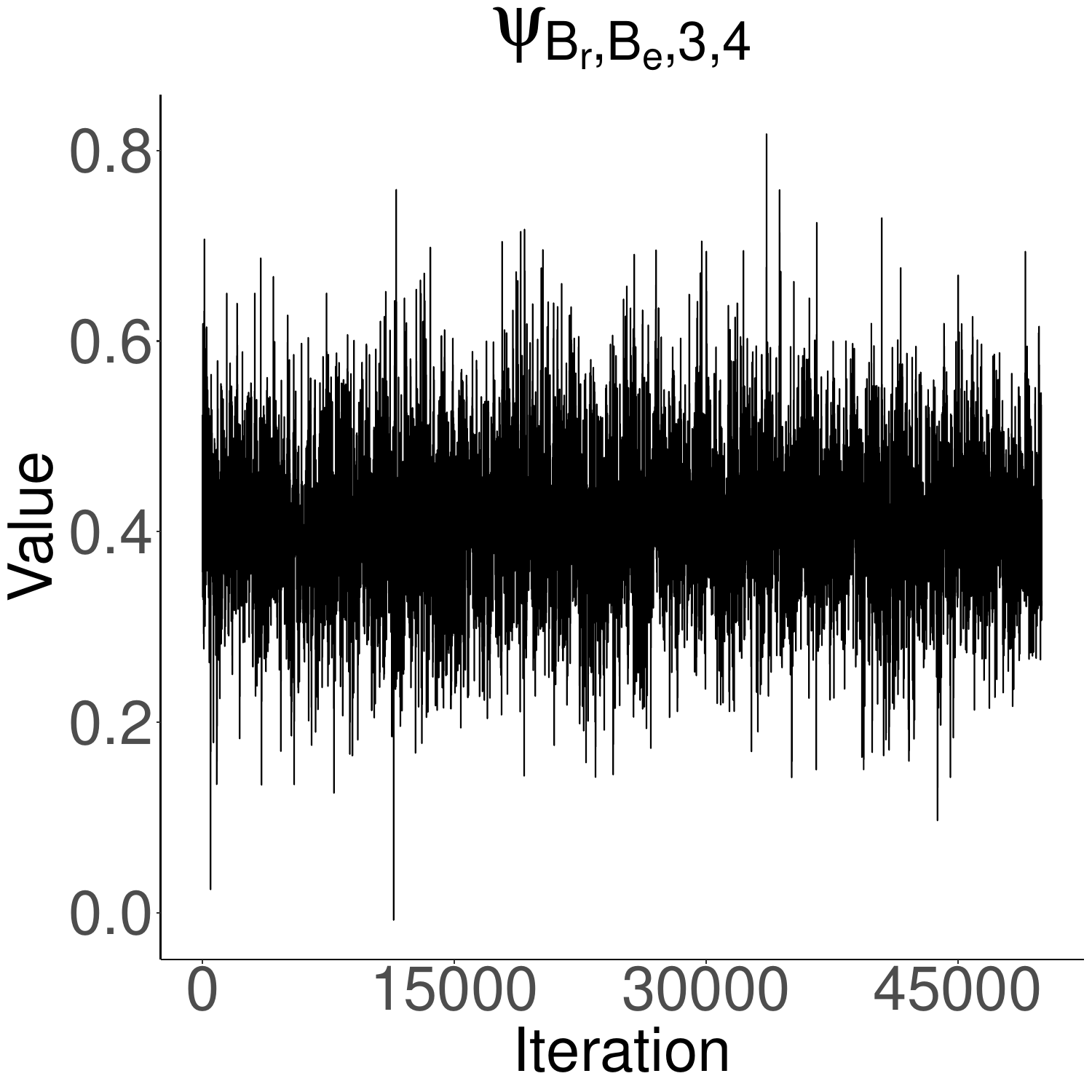}
\includegraphics[width=0.24\textwidth]{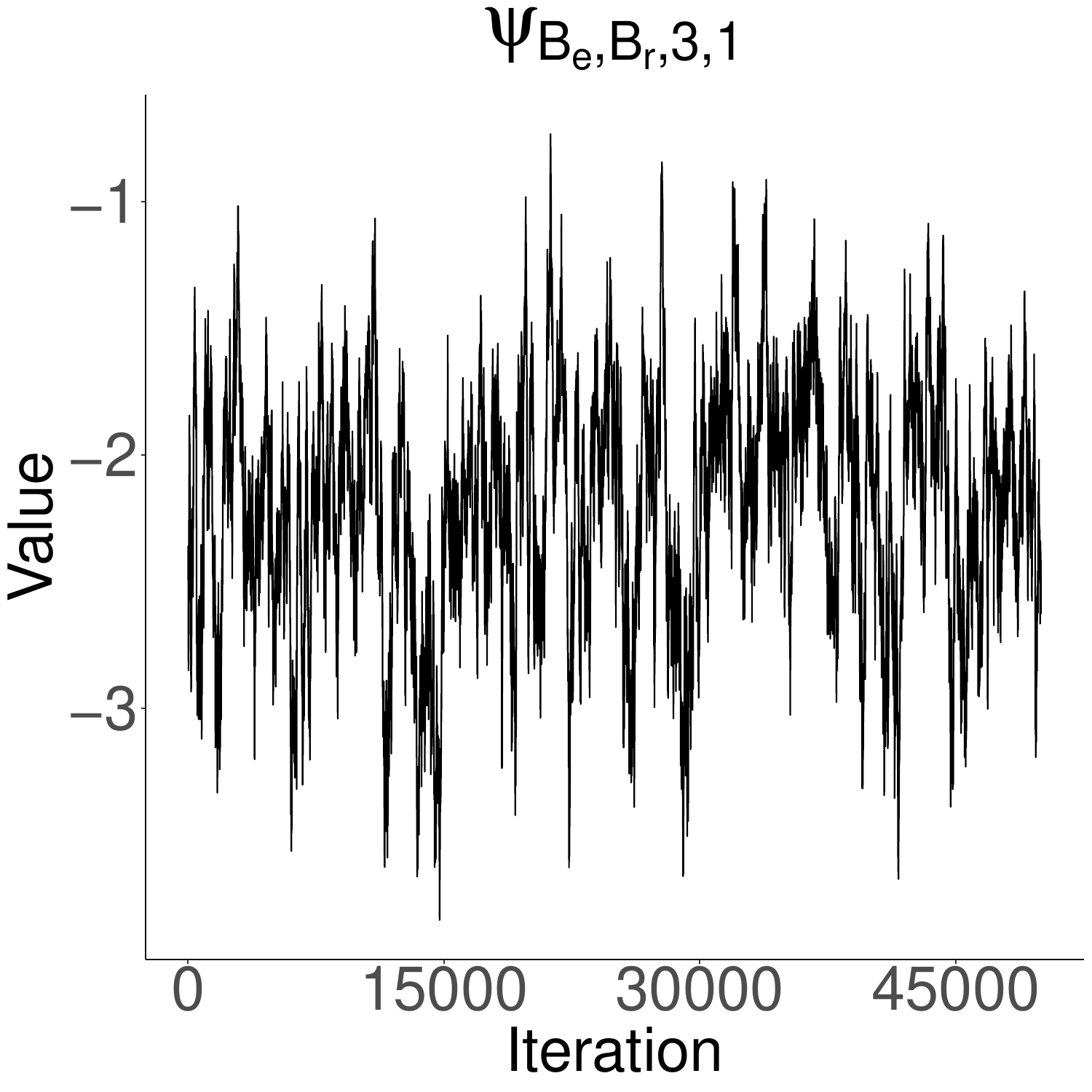}
\includegraphics[width=0.24\textwidth]{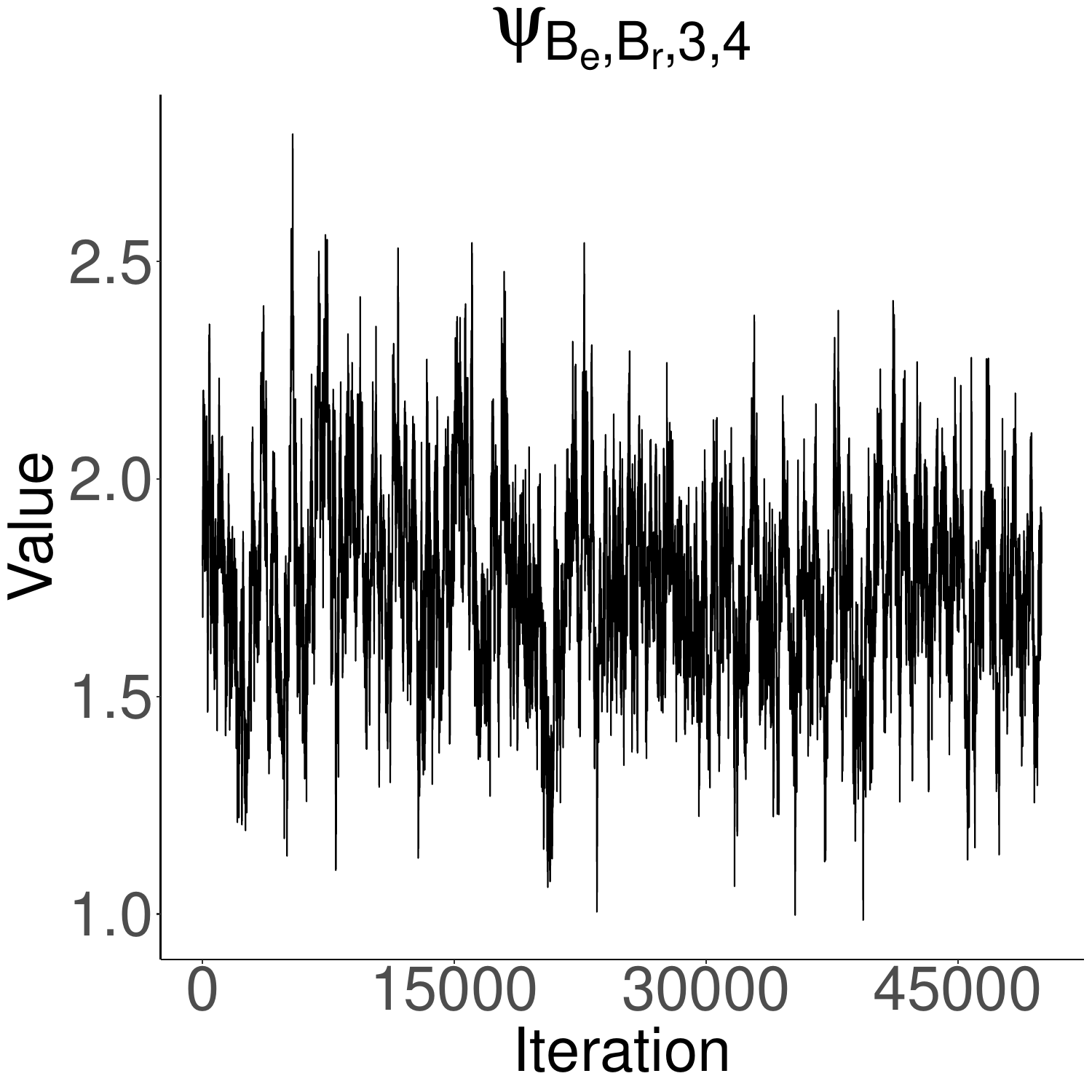}
\caption{MCMC trace plots. The notation $\bm{\Psi}_{B_r,B_e,i,j}$ found at the figure titles reflects the spatial relations from variable $i$ of beetroot on variable $j$ of Belgian endive. The trace plots contain 50000 draws from the posterior distributions of the spatial effects.}
\label{fig:diagresults}
\end{figure}

\noindent In order to adhere to the identifiability restrictions of Assumption \ref{ass:identifiability}, this application makes use of informative normal priors for the spatial effects, whereby knowledge obtained from research is translated into priors. Based on Czaban et al.\ (\citeyear{czaban2023enhancing}), who evaluated an intercropping system consisting of the same two crops as the present analysis, we expect positive effects of a crop's root length on the root length, weight and height of neighbouring crops, due to the enhanced nutrient uptake by longer roots, which might be transported to neighbouring crops. Conversely, we expect that the number of plants has a negative effect on all variables of the other crop due to the arising competition effects (Ren et al., \citeyear{ren2016planting}). The last informative priors that are imposed are normal priors with a negative mean relating to the spatial effect of Belgian endive height and weight on beetroot height (Coutinho et al., \citeyear{coutinho2017establishment}).  This results in a total of 16 informative priors, leaving 16 spatial effects on which normal-gamma priors are imposed. The informative priors have means of $1$ and $-1$ for respectively positive and negative effects, with a standard deviation of $0.01$, reflecting our strong belief in non-negligable effect sizes for the relations mentioned in the literature. We run the Gibbs sampling algorithm with a burnin of 50000, for 100000 total iterations. The estimated spatial chain graph is shown in Figure \ref{fig:toepassing}, whilst diagnostic plots for a subset of the spatial effects are given in Figure \ref{fig:diagresults}.

\noindent Even though normal-gamma priors were imposed on the unknown spatial effects, the estimated spatial chain graph is very dense, see Figure \ref{fig:toepassing}, most notably for the between-category effects for which only two effects were estimated to be 0. Conversely, the within-category effects exhibit a substantial amount of sparsity. In addition, even though the informative priors were by and large the same for the two spatial effect matrices, the estimated effects illustrate the need for asymmetry. Moving on to the interpretation of the spatial chain graph, provided that the informative priors accurately reflect the underlying model, unusual negative within-category relationships between root length and weight, as well as between root length and height, are observed. A potential explanation for these negative effects is that longer roots do not necessarily lead to an increased nutrient uptake, or a nutrient uptake that encourages plant growth, as plants already acquire more nutrients evidenced from the (imposed) positive relationship between a plant's root length and the weight/height of neighbouring crops. Therefore, whenever a plant grows longer roots, which requires resources, valuable nutrients are wasted that could have gone into additional plant growth. A second within-category effect of note is the conditional independence between the number of plants and the other variables, showing that within plots, increasing the planting density has no negative effects on plant growth, despite the increased competition between plants for resources. Whereas the between-category effects of root length on height and weight were more or less fixed by informative priors, the effect of a crop's root length on a neighbouring crop's number of plants was not. This effect was positive for the root length of beetroot on the number of Belgian endive plants, but negative for the root length of Belgian endive on the number of beetroot plants. 

In order to gain more insight into the exact nature of these synergistic and competitive effects, a second, more thorough analysis is needed, consisting of data with more variables that influence plant production, such as biological characteristics of the plant, as well as soil properties and management techniques to advance intercropping research. 

\section{Conclusion and discussion} \label{Conclusion}
In this contribution, we propose a new statistical methodology that is able to infer multivariate symmetric within-location and asymmetric between-location spatial effects. In essence, the proposed model can be seen as a fusion, and extension, of two different models: the multivariate spatial autoregressive model and the Gaussian graphical model. By exploiting the flexibility offered by the spatial weight matrices, asymmetry in spatial effects is accounted for. As this results in a highly parameterised model, identifiability restrictions are imposed on the spatial effect matrices. The inferred local and spatial effects are represented by means of a graphical framework, facilitating interpretability. Extensive simulations for a variety of network structures demonstrate that the proposed method is able to effectively reconstruct both the within- and between-location effects.

A downside of the method is the restrictiveness of the identifiability assumptions. These assumptions rely on prior knowledge of the (potential absence of) spatial relations underlying the phenomenon of interest. The question that arises is that if such prior knowledge exists, why use this method to begin with? The answer is that the proposed method provides numerical values and uncertainty pertaining to the values of these effects, which goes above and beyond typical prior knowledge. In addition, the method only requires knowledge on a subset of the spatial processes underlying any observed phenomenon. The incorporation of prior knowledge in statistical methods can lead to discoveries that are otherwise unattainable, i.e.\ the full spatial dependency process, and is part of the Bayesian philosophy of statistical inference. Moreover, researchers are not necessarily restricted to the set of identifiability restrictions proposed in this article. There exist other possible identification restrictions, which might be more suited to particular research questions or settings. On a final note, the proposed methodology challenges practitioners to translate knowledge into informative priors. An example whereby existing knowledge and statistics can be combined with statistics constitutes the usage of physiological models -- such as crop growth models -- from which part of the known physiological relations can be translated into priors and applied on a more complex spatial scenario with asymmetric relations, using the proposed method. 

One recommendation for future research is to allow for a different precision matrix for each category. In some cases the assumption of stable within-location effects might be overly restrictive. On the flip side, removing such a restriction would greatly increase the computation time of the method, making it unattractive for datasets consisting of more observations and variables. In addition, it would result in additional identifiability issues. Related to this is the absence of a left-right distinction in the spatial effects, as illustrated by the application. Such a distinction could be prove to be especially valuable for statistical analyses of intercropping systems, as the influence of weather-related variables on crops can change spatial effects, depending on the whether crops are located left or right of their neighbours, and therefore merits additional investigation. Another avenue for future research is in methods that speed up computation of the proposed method, such that data consisting of many observations and variables can be accommodated for. Computing determinants of large matrices in every iteration of the Gibbs algorithm is computationally expensive, or even prohibitive for large data. Whilst (approximate) methods have been developed for symmetric matrices, heterogeneous spatial effects, or multivariate spatial effects, no existing method substantially reduced the computation time of determinants of the form introduced by the method proposed in this article. A final recommendation is to extend the proposed method with a temporal component, as many complex processes evolve over time, which might also result in different sets of within- and between-location effects at each time point. 

\section*{Acknowledgements}
The authors would like to thank Anna Edlinger, Jan Peter van der Hoeve, Peter Bourke and Piter Bijma (Wageningen University \& Research) for sharing the data. Generation of the data was supported by the Dutch Ministry of Agriculture, Nature and Food Quality (KB44-001-001).

\bibliographystyle{Chicago}
\bibliography{library}

\newpage
\section*{Supporting Information}
Web Appendix containing proofs and derivations referenced in Section 2, additional simulation results, computation time evaluation, an illustration of convergence of the Gibbs algorithm a summary of the graphical horseshoe method and the R code implementing the method are available with this paper at the Biometrics website on Wiley Online Library. The R package SAGM is available at \url{https://CRAN.R-project.org/package=SAGM}.

\section*{Data distribution}
Here we derive the expected value and variance of $\bm{X}$, where
\begin{equation*}
\begin{gathered}
    \mathbb{E}[\text{vec}(\bm{X})] = \mathbb{E}\left[\bm{R}(\bm{\Psi})\text{vec}(\bm{E})\right]\\
      = \bm{R}(\bm{\Psi})\mathbb{E}[\text{vec}(\bm{E})]\\
       = \bm{0}.
    \end{gathered}
\end{equation*}
and 
\begin{equation*} 
\begin{gathered}
        \text{Var}[\text{vec}(\bm{X})] = \mathbb{E}[\text{vec}(\bm{X})\text{vec}(\bm{X})^T]\\
        = \mathbb{E}\left\{\left[\bm{R}(\bm{\Psi})^{-1}\text{vec}(\bm{E})\right]\left[\bm{R}(\bm{\Psi})^{-1}\text{vec}(\bm{E})\right]^T\right\}\\
        = \bm{R}(\bm{\Psi})^{-1}( \bm{\Sigma}_E \otimes \bm{I}_n)\bm{R}(\bm{\Psi})^{-T}.
\end{gathered}
\end{equation*}
Consequently, $\text{vec}(\bm{X})$ is distributed as $\text{vec}(\bm{X}) \sim N_{np}\left(\bm{0}, \bm{R}(\bm{\Psi})^{-1}( \bm{\Sigma}_E \otimes \bm{I}_n)\bm{R}(\bm{\Psi})^{-T}\right)$, shortened as $\text{vec}(\bm{X}) \sim N_{np}(\bm{0},  \bm{\Sigma}_{X})$. 

\section*{Proofs}
Here we provide the proofs for the theorems found in Section 2.
\\
\\
\noindent Proof of Theorem 1
\begin{proof}
The vectors consisting of spatially filtered data $\bm{E}_i$ are independently distributed as $N(\bm{0}, \bm{\Theta}_E^{-1}), i = 1,\ldots,n$, where information pertaining to within-category relations for any category $c$ can be found in $ \bm{\Theta}_E$. This property of independence conditional on a separating set is known as the global Markov property. Lauritzen (\citeyear{lauritzen1996graphical}) provides a proof that any data arising from a multivariate normal distribution satisfy the global Markov property on its graph. 
\end{proof}

\noindent Proof of Theorem 2
\begin{proof}
Suppose for simplicity, and without loss of generality, that $\bm{X}_{c_1,c_2} \cup \bm{X}_{c_3,c_4} = \bm{X}$ and that $A_{c_1,c_2}$ and $B_{c_3,c_4}$ contain an equal number of observations $n$ for a total number of $2n$ observations in $\bm{X}$. We can represent the spatial weight matrices for the observations pertaining to $A_{c_1,c_2}$ as $\overline{\bm{W}}_{c_1,c_2} = \begin{pmatrix}
\bm{W}_{c_1,c_2} & \bm{O}_n\\
\bm{O}_n & \bm{O}_n
\end{pmatrix}$ and $\overline{\bm{W}}_{c_2,c_1} = \begin{pmatrix}
\bm{W}_{c_2,c_1} & \bm{O}_n\\
\bm{O}_n & \bm{O}_n
\end{pmatrix}$ and those of $B_{c_3,c_4}$ as $\overline{\bm{W}}_{c_3,c_4} = \begin{pmatrix}
\bm{O}_n & \bm{O}_n\\
\bm{O}_n & \bm{W}_{c_3,c_4} 
\end{pmatrix}$ and $\overline{\bm{W}}_{c_4,c_3} = \begin{pmatrix}
\bm{O}_n & \bm{O}_n\\
\bm{O}_n & \bm{W}_{c_4,c_3} 
\end{pmatrix}$, with $\bm{O}_n$ being a $n \times n$ zero matrix, where the various $\bm{W} \in [0,1]^{n \times n}$ are sub-neighbour matrices only containing locations with the categories stated in the subscript, and with the $\overline{\bm{W}} \in [0,1]^{2n \times 2n}$. Due to the block structure of the weight matrices, the spatial filter matrix $\bf{R}(\bf{\Psi})$ becomes separable for the $\bm{\Psi}_k$, such that
\begin{equation*}
    \begin{gathered}
        \begin{pmatrix}
    \left(\bm{I}_{2np} - \bm{\Psi}_{c_1,c_2}^T \otimes \overline{\bm{W}}_{c_1,c_2} - \bm{\Psi}_{c_2,c_1}^T \otimes \overline{\bm{W}}_{c_2,c_1}\right)_{[c_1,c_2]}^{-1}\\
    \left(\bm{I}_{2np} - \bm{\Psi}_{c_3,c_4}^T \otimes \overline{\bm{W}}_{c_3,c_4} - \bm{\Psi}_{c_4,c_3}^T \otimes \overline{\bm{W}}_{c_4,c_3}\right)_{[c_3,c_4]}^{-1}
\end{pmatrix}\\
= \left(\bm{I}_{2np} - \bm{\Psi}_{c_1,c_2}^T \otimes \overline{\bm{W}}_{c_1,c_2} - \bm{\Psi}_{c_2,c_1}^T \otimes \overline{\bm{W}}_{c_2,c_1} - \bm{\Psi}_{c_3,c_4}^T \otimes \overline{\bm{W}}_{c_3,c_4} - \bm{\Psi}_{c_4,c_3}^T \otimes \overline{\bm{W}}_{c_4,c_3}\right)^{-1},
    \end{gathered}
\end{equation*}
where $[c_1,c_2]$ indicates the rows that pertain to observations of categories $c_1$ and $c_2$. Above result shows the independence of the data generating process of combinations $c_1,c_2$ and $c_3,c_4$ and completes the proof.
\end{proof}

\noindent Proof of Corollary 1
\begin{proof}
As the spatial effect matrices $\bm{\Psi}_{c_2,c_1}$ and $\bm{\Psi}_{c_1,c_2}$ both equal $\bm{O}_p$, we have that the data generating process reduces to $\text{vec}(\bm{X}) = \text{vec}(\bm{E})$, indicating that the data depend only on the within-location effects, thereby confirming the independence between any subsets of $\bm{X}_{c_1}$ and $\bm{X}_{c_2}$.
\end{proof}

\section*{On the non-identifiability of the spatial autoregressive graphical model}
The proposed method is not able to identify the $\bm{\Psi}_k$. By identify, we refer to likelihood-identifiability, that is, the value of the likelihood is different for different parameter values. The non-identifiability for the $\bm{\Psi}_k$ of the spatial autoregressive graphical model is illustrated by means of an example. Consider again the toy example provided in the main section, in which $n = 5$ and where we have the following weight matrices
\begin{equation*}
    \bm{W}_{c_2,c_1} = \begin{pmatrix}
0 & 1 & 0 & 0 & 0\\
0 & 0 & 0 & 0 & 0\\
0 & 0.5 & 0 & 0.5 & 0\\
0 & 0 & 0 & 0 & 0\\
0 & 0 & 0 & 1 & 0
\end{pmatrix} \quad \bm{W}_{c_1,c_2} = \begin{pmatrix}
0 & 0 & 0 & 0 & 0\\
0.5 & 0 & 0.5 & 0 & 0\\
0 & 0 & 0 & 0 & 0\\
0 & 0 & 0.5 & 0 & 0.5\\
0 & 0 & 0 & 0 & 0
\end{pmatrix}.
\end{equation*}
Moreover, the following data on $p = 2$ variables are obtained
\begin{equation*}
    \bm{X} = \begin{pmatrix}
    	1.6154947 & -1.2923845\\
	-0.6613634 &  1.4513324\\
	0.2156944 &  0.5118063\\
	-0.5619332 & -0.8025372\\
	0.2537716 & -2.4355581\\
\end{pmatrix}, 
\end{equation*}
where the underlying parameters are
\begin{equation*}
    \bm{\Psi}_{c_1,c_2} = \begin{pmatrix}
-0.1 & -0.2\\
0 & 0.4 \end{pmatrix} \quad \bm{\Psi}_{c_2,c_1} = \begin{pmatrix}
0 & 0.4\\
-0.4 & 0.4
\end{pmatrix},
\quad  \bm{\Theta}_E = \begin{pmatrix}
 1 & 0.1\\
0.1 &  1
\end{pmatrix}.
\end{equation*}
Computing the log likelihood (see Equation (4) in the main text) results in a value of -7.165605. However, when fixing $\hat{\bm{\Theta}}_{E}$ to $\bm{\Theta}_E$, we obtain the same value for the log likelihood when the following autoregressive parameters are used
\begin{equation*}
    \bm{\Psi}_{c_1,c_2} = \begin{pmatrix}
	-0.5552457 & -0.02928151\\
	0.1116245 & -0.56307484 \end{pmatrix} \quad \bm{\Psi}_{c_2,c_1} = \begin{pmatrix}
	0.1067707 & -0.1474186\\
	-0.5431019 & 0.7314883
\end{pmatrix},
\end{equation*}
showing that the model is not identifiable for the $\bm{\Psi}_k$.

\section*{On the separate estimation of spatial effect matrices}
\noindent Separate estimation of the spatial effect matrices is impossible, as we have that 

\begin{equation*}
\begin{gathered}
\begin{pmatrix}
    \left(\bm{I}_{np} - \bm{\Psi}_{c_1,c_2}^T \otimes \bm{W}_{c_1,c_2}\right)_{[c_2]}^{-1}\\
    \left(\bm{I}_{np} - \bm{\Psi}_{c_2,c_1}^T \otimes \bm{W}_{c_2,c_1}\right)_{[c_1]}^{-1}
\end{pmatrix}\\
 \neq \left(\bm{I}_{np} - \bm{\Psi}_{c_1,c_2}^T \otimes \bm{W}_{c_1,c_2} - \bm{\Psi}_{c_2,c_1}^T \otimes \bm{W}_{c_2,c_1}\right)^{-1}\\
 \neq \begin{pmatrix}
    \left(\bm{I}_{np} - \bm{\Psi}_{c_1,c_2}^T \otimes \bm{W}_{c_1,c_2}\right)_{[c_1]}^{-1}\\
    \left(\bm{I}_{np} - \bm{\Psi}_{c_2,c_1}^T \otimes \bm{W}_{c_2,c_1}\right)_{[c_2]}^{-1}
\end{pmatrix},
\end{gathered}
\end{equation*}
for $c_2 \in \mathcal{N}_{c_1}$, due to the interdependence between the two categories. Note that the $\begin{pmatrix}
    \cdot\\
    \cdot
\end{pmatrix}$ 
 does not represent explicit matrix stacking as in the proof of Theorem 2, but instead represents a matrix where we stack the rows of the spatial filter matrices according to their location in the intercropping design. 

\section*{Graphical Horseshoe}
The graphical horseshoe (Li et al., \citeyear{li2019graphical}) is briefly recapitulated. We make use of an adapted graphical horseshoe, because of the fact that we are not only interested in $\bm{\Theta}_E$, but also in the $\bm{\Psi}_k$. In essence, this only changes the computation of $\bm{S}$, which is computed as $\bm{S} = \bm{X}^{T}\bm{X}$ in the original version of the algorithm. At every iteration in the Gibbs sampling algorithm, a graphical horseshoe step is included to update the estimates of $\bm{\Theta}_E$, which in turn is used to update the estimates for the $\bm{\Psi}_k$.
 
\begin{algorithm}
\caption{The adapted Graphical Horseshoe Sampler}\label{alg:gibbs}
    \hspace*{\algorithmicindent} \textbf{Input:} $\bm{X}$, $\bm{W}_1$ and $\bm{W}_2$\\
    \hspace*{\algorithmicindent} \textbf{Output:} $\hat{\bm{\Theta}}_E$
  \begin{algorithmic}[1]
        \STATE Initialise $\bm{\Theta}_E = \bm{I}_{p}$,  $\bm{\Sigma}_E = \bm{I}_{p}$, $\bm{\Lambda}$ = $\mathbbm{1}_p$, $N = \mathbbm{1}_p$, $\xi = 1, \zeta = 1$, $\bm{\Psi}_1 = \bm{\Psi}_2 = \bm{O}_p + \epsilon$, for small $\epsilon > 0$.
        \STATE Compute $\bm{S} = \left(\bm{X}- \sum_{k = 1}^{2} \bm{W}_{k}\bm{X}\bm{\Psi}_k\right)^T \left(\bm{X}-\sum_{k = 1}^{2} \bm{W}_{k}\bm{X}\bm{\Psi}_k\right)$
    \FOR{$i = 1$ to $p$}
        \STATE Sample $\gamma \sim G(\frac{n}{2}+1, \frac{s_{i,i}}{2})$
        \STATE Set $\bm{\Theta}^{-1}_{E,-i,-i} = \bm{\Sigma}_{E,-i,-i} - \frac{\bm{\sigma}_{E,-i,i}\bm{\sigma}_{E,-i,i}^{T}}{\sigma_{E,i,i}}$
        \STATE Set $C = \left(s_{ii}\bm{\Theta}^{-1}_{E,-i,-i} + \text{diag}\left(\bm{\lambda_{-i,i}}\xi^2\right)^{-1}\right)^{-1}$
        \STATE Sample $\bm{\beta} \sim N(-C\bm{s}_{-i,i},C)$
        \STATE Set $\bm{\theta}_{E,-i,i} = \bm{\beta}, \theta_{ii} = \gamma + \bm{\beta}^{T}\bm{\Theta}^{-1}_{E,-i,-i}\bm{\beta}$
        \STATE Sample $\bm{\lambda}_{-i,i} \sim IG(1,\frac{1}{\bm{\nu}_{-i,i}} + \frac{\bm{\theta}^2_{-i,i}}{2\xi^2})$
        \STATE Sample $\bm{\nu}_{-i,i} \sim IG(1, 1+ \frac{1}{\bm{\lambda}_{-i,i}})$
        \STATE Set $\bm{\Sigma}_{E, -i, -i} = \bm{\Theta}_{E, -i, -i}^{-1} + \frac{(\bm{\Theta}_{E, -i, -i}\bm{\beta})(\bm{\Theta}_{E, -i, -i}^{-1}\bm{\beta})^{T}}{\gamma}, \bm{\sigma}_{E,-i,i} = \frac{-(\bm{\Theta}_{E, -i, -i}^{-1}\bm{\beta})}{\gamma}, \sigma_{E,i,i} = \frac{1}{\gamma}$
    \ENDFOR
    \STATE Sample $\xi^2 \sim IG\left(\frac{(\binom{p}{2} + 1)}{2}, \frac{1}{\zeta} + \frac{\sum_{i<j}\theta_{E,i,j}^2}{2\lambda_{E,i,j}^2}\right)$
    \STATE Sample $\zeta \sim IG\left(1, 1+\frac{1}{\xi^2}\right)$
  \end{algorithmic}
\end{algorithm}

\section*{MCMC convergence}

\begin{figure}[H]
\centering
\includegraphics[width=0.32\textwidth]{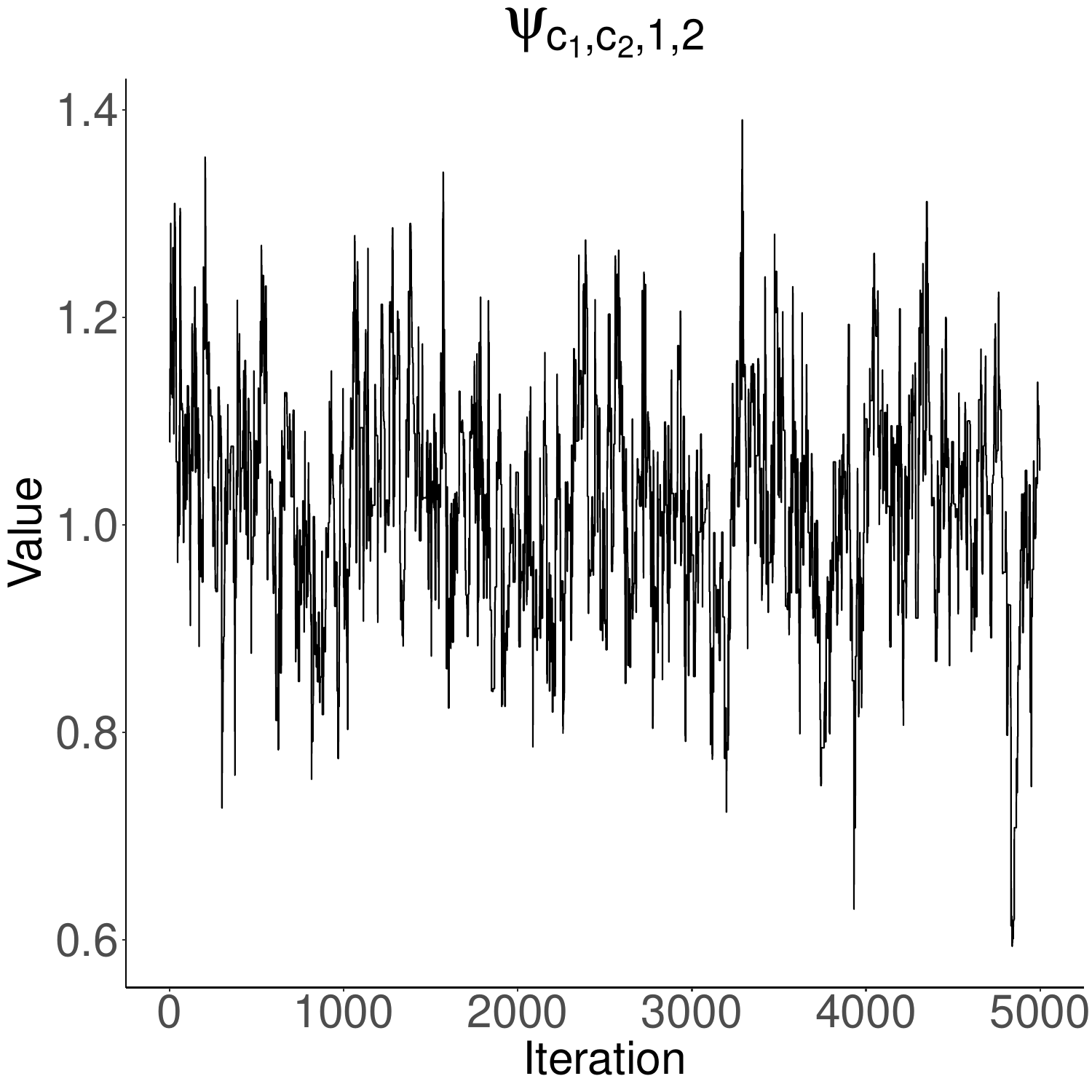}
\includegraphics[width=0.32\textwidth]{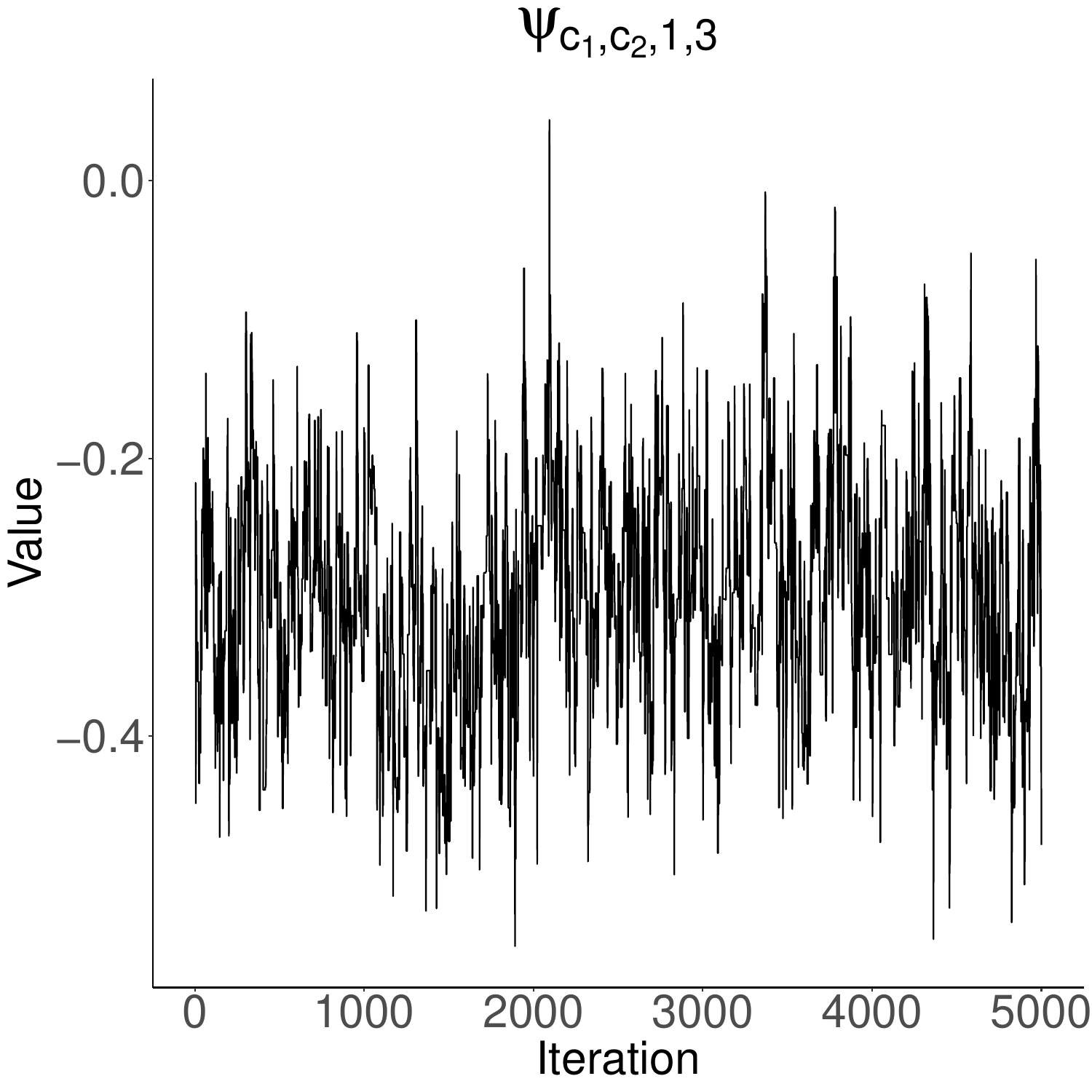}
\includegraphics[width=0.32\textwidth]{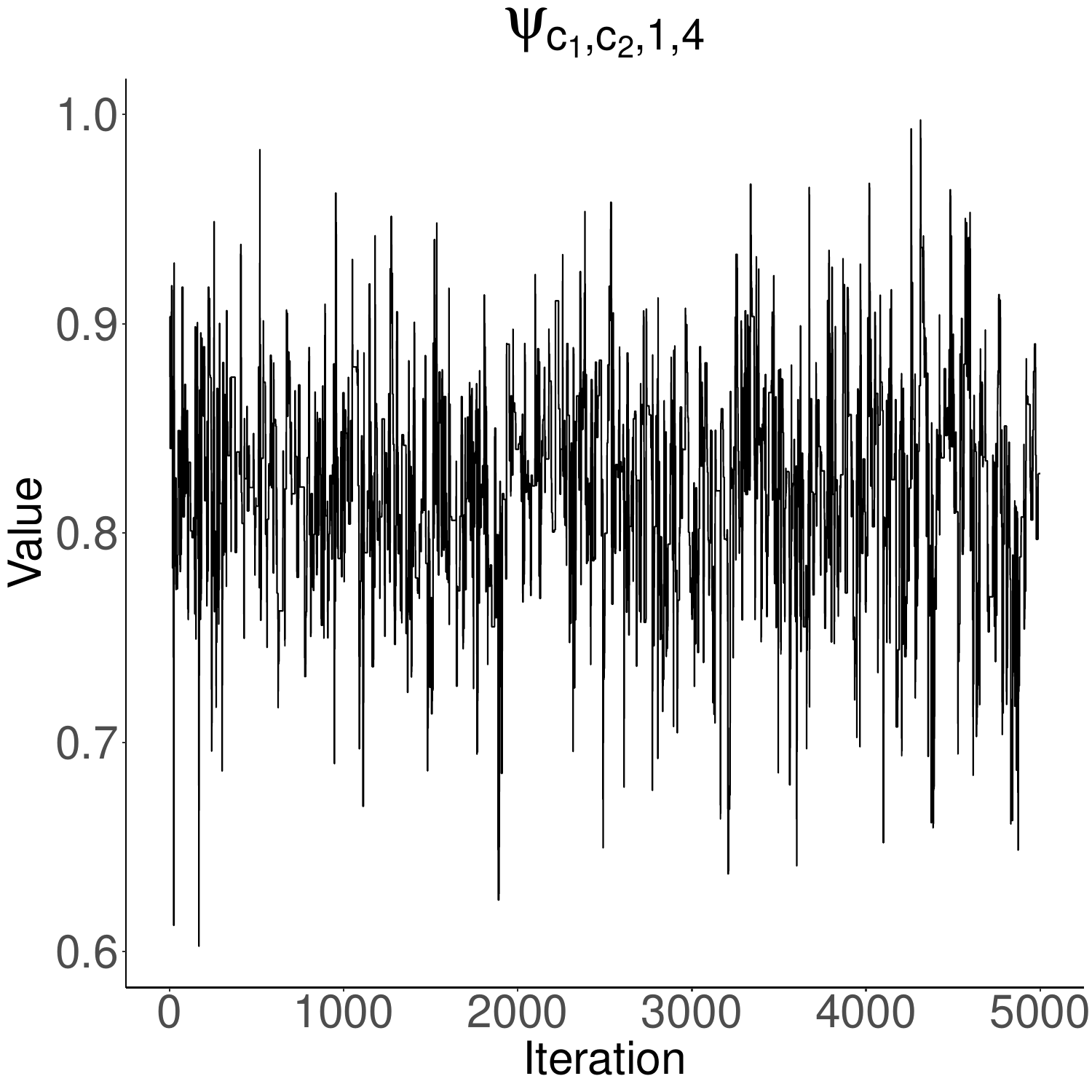}\\
\includegraphics[width=0.32\textwidth]{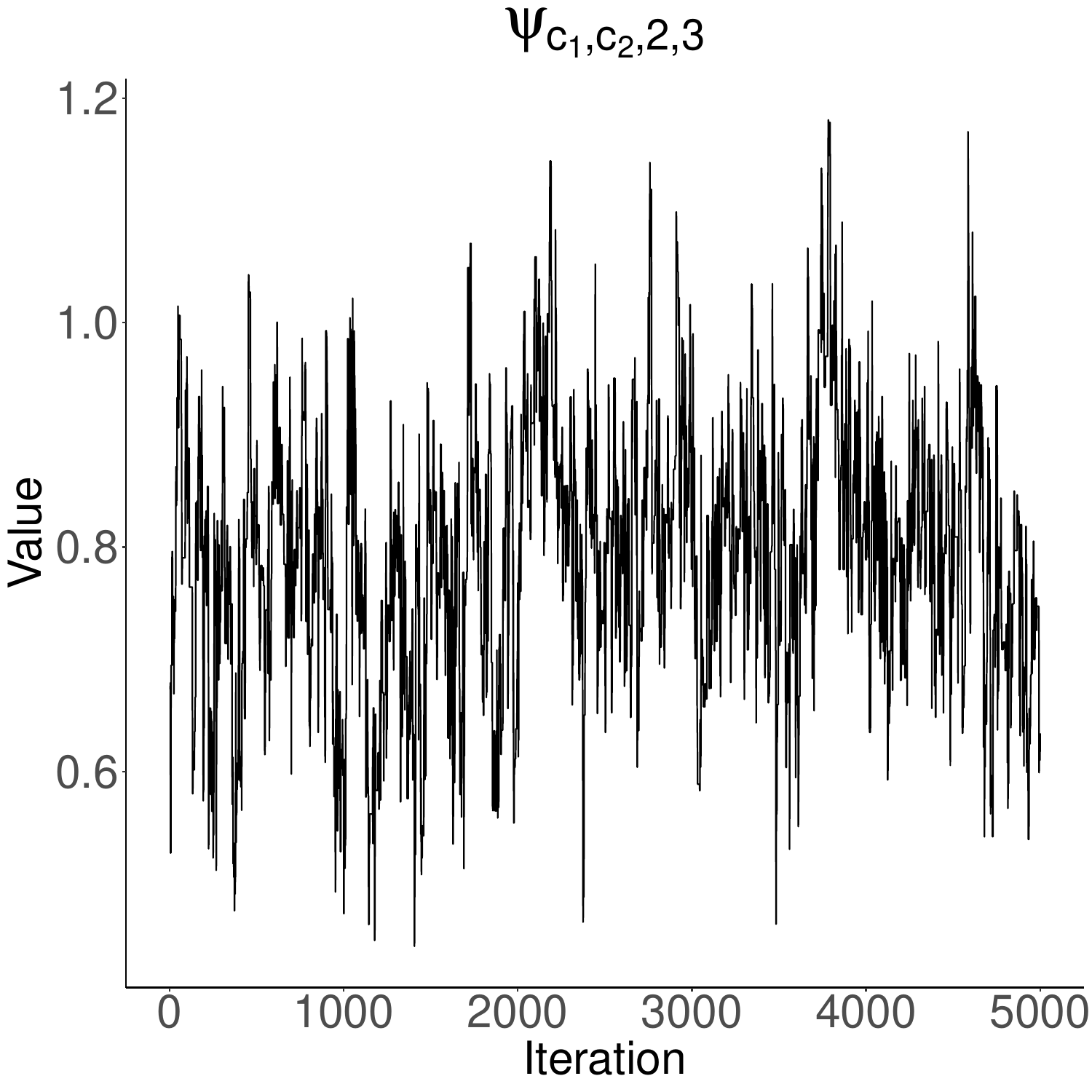}
\includegraphics[width=0.32\textwidth]{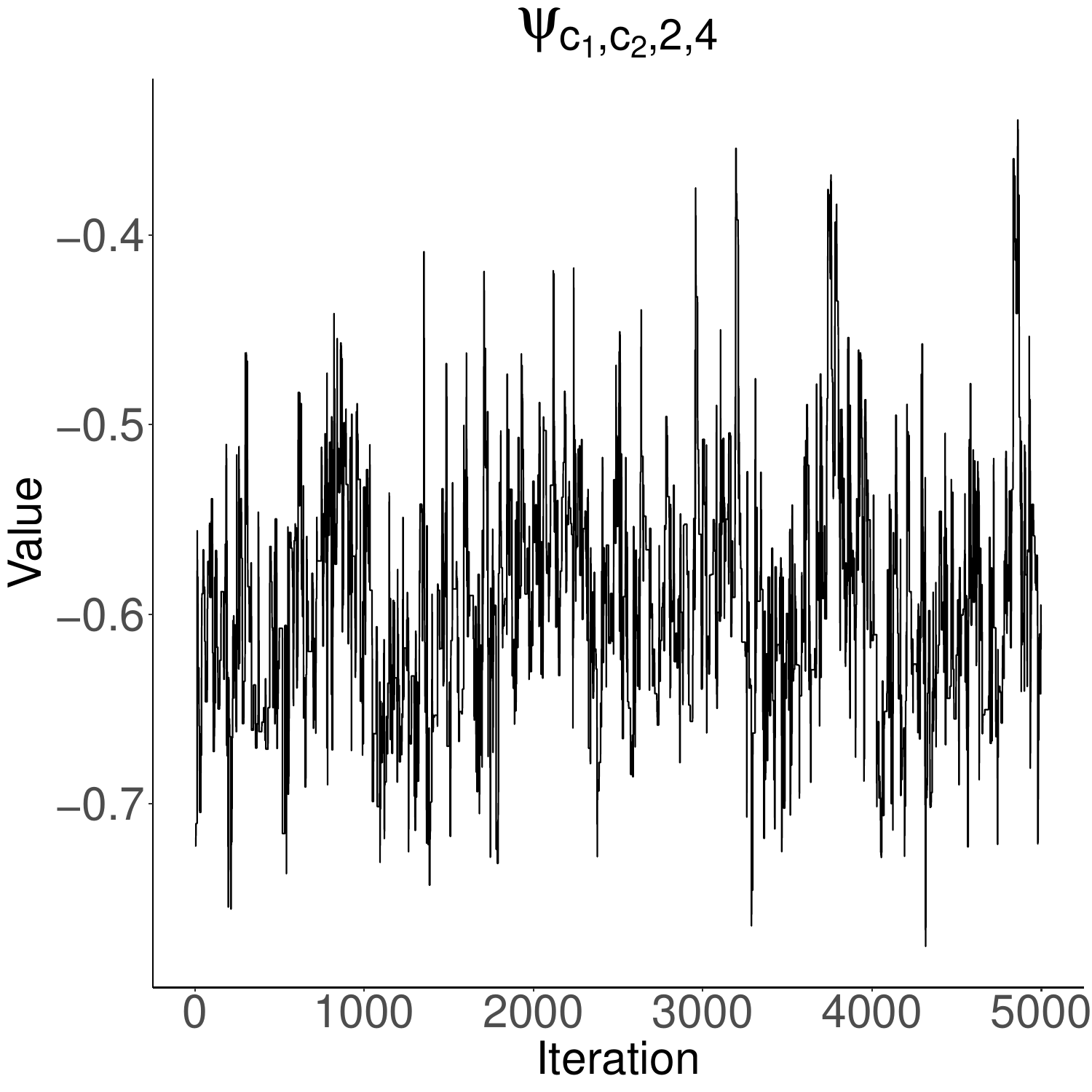}
\includegraphics[width=0.32\textwidth]{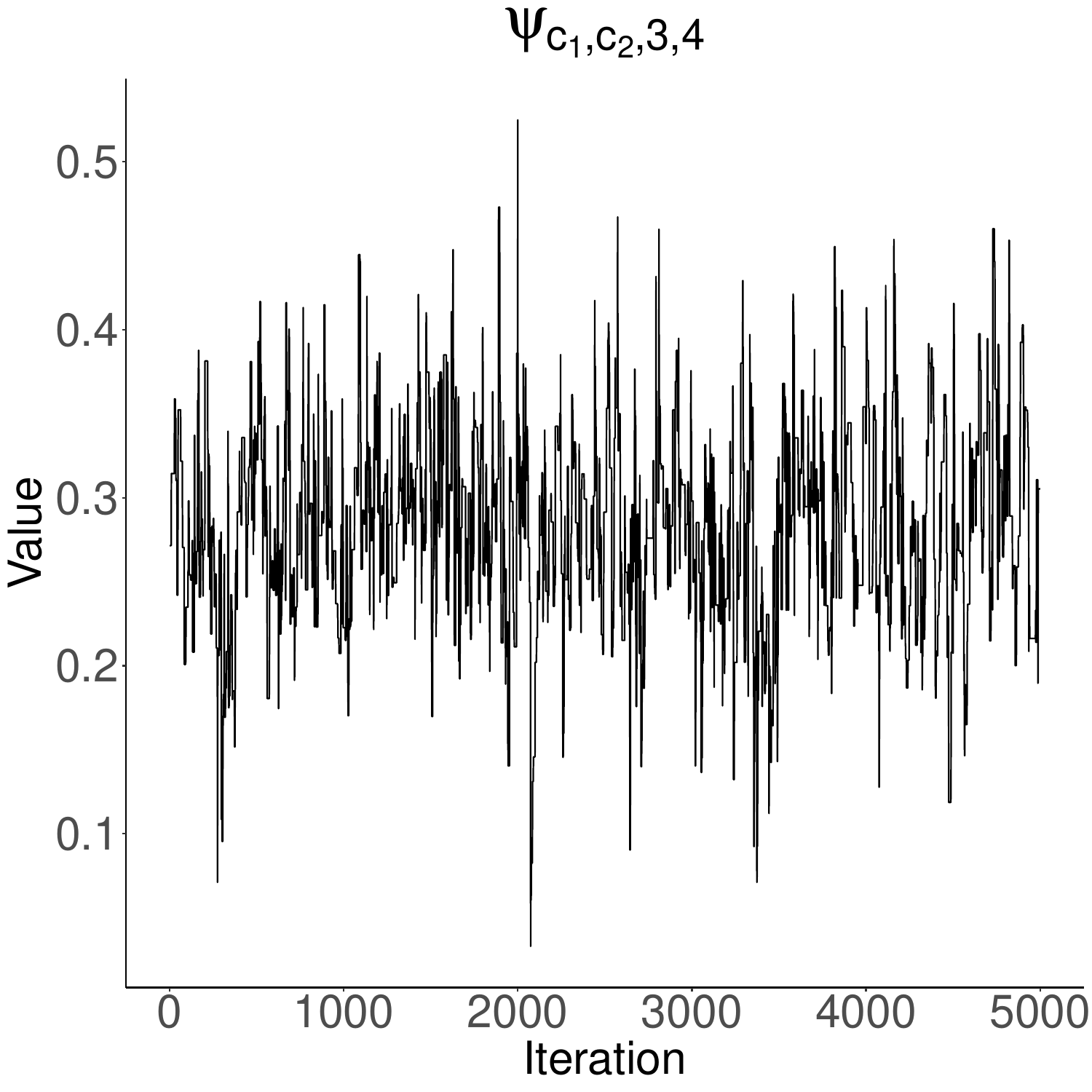}
\caption{MCMC trace plots and posterior distributions for the upper triangular elements of $\bm{\Psi}_{c_1,c_2}$ obtained in the first random seed from the simulation using settings $n = 25$, $p = 4$, a random network for $\bm{\Theta}_E$ and normal priors.}
\label{fig:MCMC_conv}
\end{figure}

To illustrate convergence of the Gibbs method, we use simulated data and evaluate trace- and autocorrelation plots of the Gibbs samples. Both plots are based on the post-burn-in samples. The data used for this study consists of a random network for $\bm{\Theta}_E$, with $n = 25$, $p = 4$ and (known) symmetry for the $\bm{\Psi}_k$. The spatial autoregressive model was fitted using normal priors.  When running the Gibbs method for the simulation studies, we used a single chain with a burnin of 5000, with 5000 iterations post burnin. We show only the plots for the first 6 variables in Figure \ref{fig:MCMC_conv}.

\section*{Computation time}
We showcase the computation time of the proposed method, where the data is simulated in the same way as in Section 4 of the article, except that in this case, we fix $p = 10$, let $n \in \{5,10,25,50,100\}$ and only evaluate the computation time of the model on random networks. Nevertheless, the computation time for other network types is very similar. The simulations were conducted using an AMD Ryzen 5 2600 3.4Ghz processor with 16GB of RAM.

\begin{table}[H]
\resizebox{\textwidth}{!}{%
\begin{adjustbox}{}
\centering
  \begin{threeparttable}
  \caption{Computation time (in minutes) for the random network, using both priors for both symmetric and triangular parameter restrictions. The computation time is averaged across 20 fitted models for each parameter combination and rounded to 2 decimals. Standard errors are provided between parentheses.}
  \label{tab:simstudy}
     \begin{tabular}{l| c | c | c | c}
        \toprule
        \midrule
        \multicolumn{1}{c}{}  & \multicolumn{2}{|c|}{Normal prior}  & \multicolumn{2}{c}{Normal-gamma prior}\\
        \midrule
        \multicolumn{1}{c|}{$n, p$} & \multicolumn{1}{c|}{Symmetric restriction} & \multicolumn{1}{c|}{Triangular restriction}  & \multicolumn{1}{c|}{Symmetric restriction} & \multicolumn{1}{c}{Triangular restriction}\\
        \midrule
$5, 10$ & 2.51 (0.03) & 5.45 (0.02) & 3.21 (0.00) & 5.28 (0.00)\\
$10, 10$ & 6.42 (0.03) & 10.1 (0.01) & 5.86 (0.01) & 9.91 (0.02)\\
$25, 10$ & 47.03 (0.23) & 78.12 (0.09) & 41.94 (0.00) & 78.65 (0.05)\\
$50, 10$  & 323.79 (0.83) & 596.87 (4.13) & 357.94 (4.32) & 544.52 (3.16)\\
$100, 10$ & 2417.47 (2.06) & 3665.11 (10.54) & 2362.18 (3.92) & 3490.60 (8.64)\\
\midrule
\bottomrule
     \end{tabular}
  \end{threeparttable}
    \end{adjustbox}}
\end{table}

\end{document}